%% file: main.tex
\documentclass[11pt,a4paper]{article}
\pdfoutput=1
\usepackage{jheppub}
\usepackage{amsmath,amsfonts,amssymb,bbm}
\usepackage{hyperref, csquotes}
\usepackage{graphicx, color}
\usepackage{tikz,pgf}
\include{fdiagrams}

\usepackage{comment}

\newcommand{\qmark}[1]{``#1''}
\renewcommand{\eqref}[1]{Eq.\,(\ref{#1})}
\renewcommand\[{\begin{equation}}
\renewcommand\]{\end{equation}}

\def\i{\mathrm i}
\def\d{\mathrm{d}}
\newcommand{\one}{\mathbb I}
\newcommand{\R}{\mathbb R}

\newcommand{\Tr}{\mathrm{Tr}}
\newcommand{\tr}{\mathrm{tr}}
\newcommand*\diff{\mathop{}\!\mathrm{d}}

\newcommand{\std}{{d_{\text{l}}}}			    
\newcommand{\gm}{G}                 
\newcommand{\gd}{{d_{\textsc{g}}}}	
\newcommand{\cas}{\textrm{Cas}}     
\newcommand{\cs}{a_G}             	
\newcommand{\rv}{a}                 
\newcommand{\rvp}{\tilde{a}}        
\newcommand{\rk}{r}         		
\newcommand{\gf}{\Phi}       		
\newcommand{\gfb}{\Phi}
\newcommand{\cgf}{\Phi_0}       	
\newcommand{\pgf}{\delta\Phi}       

\newcommand{\ff}{\phi}       		
\newcommand{\vff}{\pmb{\phi}}       
\newcommand{\fm}{k}                 
\newcommand{\vfm}{\pmb{k}} 
\newcommand{\rep}{j}
\newcommand{\vrep}{{\pmb{\rep}}}
\renewcommand{\ell}{c}

\newcommand{\vg}{\pmb{g}}
\newcommand{\vh}{\pmb{h}}

\newcommand{\kin}{\mathcal{K}}                
\newcommand{\nlop}{\mathcal{X}}
\newcommand{\nloph}{\hat{\mathcal{X}}}
\newcommand{\nlc}{\mathcal{X}^{(\gamma)}}
\newcommand{\V}{\mathcal{V}}
\newcommand{\nv}{V}
\newcommand{\ooi}{V}
\newcommand{\ct}{\tilde{\lambda}}
\newcommand{\nei}{E}

\newcommand{\ed}{{d_{\textrm{nl}}}}
\newcommand{\cc}{\alpha}                
\newcommand{\xiloc}{\xi_{\text{l}}}      
\newcommand{\xinloc}{\xi_{\text{nl}}}      

\begin{document}

\title{Phase transitions in tensorial group field theories: Landau-Ginzburg analysis of models with both local and non-local degrees of freedom}
\author[a,b,c]{Luca Marchetti,}
\author[a]{Daniele Oriti,}
\author[a]{Andreas G. A. Pithis,}
\author[d,e]{Johannes Th\"urigen}

\emailAdd{luca.marchetti@phd.unipi.it}
\emailAdd{daniele.oriti@physik.lmu.de}
\emailAdd{andreas.pithis@physik.lmu.de}
\emailAdd{johannes.thuerigen@uni-muenster.de}

\affiliation[a]{Arnold Sommerfeld Center for Theoretical Physics,\\ Ludwig-Maximilians-Universit\"at München \\ Theresienstrasse 37, 80333 M\"unchen, Germany, EU}
\affiliation[b]{Università di Pisa,\\Lungarno Antonio Pacinotti 43, 56126 Pisa, Italy, EU}
\affiliation[c]{Istituto Nazionale di Fisica Nucleare sez. Pisa,\\Largo Bruno Pontecorvo 3, 56127 Pisa, Italy, EU}
\affiliation[d]{Mathematisches Institut der Westf\"alischen Wilhelms-Universit\"at M\"unster\\ Einsteinstr. 62, 48149 M\"unster, Germany, EU}
\affiliation[e]{Institut f\"ur Physik/Institut f\"ur Mathematik der Humboldt-Universit\"at zu Berlin\\
Unter den Linden 6, 10099 Berlin, Germany, EU
}

\begin{abstract}
{In the tensorial group field theory approach to quantum gravity, the theory is based on discrete building blocks and continuum spacetime is expected to emerge from their collective dynamics, possibly at criticality, via a phase transition.
On a compact group of fixed volume this can be expected to be only possible in a large-volume or thermodynamic limit.
Here we show how phase transitions are possible in TGFTs in two cases: a) considering the non-local group degrees of freedom on a non-compact Lie group instead of a compact one (or taking a large-volume limit of a compact group); b) in models including $\mathbb{R}$-valued local degrees of freedom (that can be interpreted as discrete scalar fields, often used in this context to provide a matter reference frame).
After adapting the Landau-Ginzburg approach to this setting of mixed local/non-local degrees of freedom, we determine the critical dimension beyond which there is a Gaussian fixed point and a continuous phase transition which can be described by mean-field theory. 
This is an important step towards the realization of a phase transition to continuum spacetime in realistic TGFT models for quantum gravity.}
\end{abstract}

\setcounter{tocdepth}{2}

\maketitle

\section{Introduction}

One of the key challenges for quantum gravity approaches based on discrete structures is the recovery of the usual description of macroscopic physics in terms of continuum spacetime and geometry. A rather general expectation is that the emergence of such continuum description requires some form of critical behaviour and an appropriate phase transition~\cite{Oriti:2013jga}. This is the case, with a number of technical and conceptual differences specific to the various formalisms, in approaches like dynamical triangulations~\cite{CDT}, quantum Regge calculus~\cite{Williams:2007up}, (canonical and covariant) loop quantum gravity~\cite{Ashtekar:2004eh,Rovelli:2011tk,Conrady:2010kc,Conrady:2010vx,Perez:2012wv}, causal set theory~\cite{Surya:2019ndm} and tensorial group field theories (TGFT), both in their purely combinatorial version given by tensor models~\cite{GurauBook} (closely related to dynamical triangulations) and in the quantum geometric (\lq group field theory\rq) models~\cite{Oriti:2012wt,Krajewski:2012wm,Carrozza:2013oiy} (closely  related to loop quantum gravity).

Tensorial group field theories\footnote{We adopt TGFT as a general term for the whole framework, and do not imply that the models we consider necessarily (but they may) have tensor symmetries (e.g. invariance under some unitary group acting on the tensor indices), which is often associated with the label \enquote{tensorial}, nor that they have specific group symmetries or depend on specific group-theoretic constructions (but they may), as often associated with the label \enquote{group}. The terms \enquote{tensorial} and \enquote{group} are justified simply by having a tensor field defined on a group manifold.} are combinatorially non-local quantum and statistical field theories of geometric degrees of freedom. As such, they can be seen as a way to extend the success of matrix models for $2d$-gravity~\cite{DiFrancesco:1993cyw} to higher dimensions. The fundamental fields live on $\rk$ copies of a Lie group (or dual spaces, related by generalization of the Fourier transform), with these group-theoretic data labelling the quanta of the field, corresponding to $(\rk-1)$-simplices and the (boundary) states of the theory correspond to gluings of such $(\rk-1)$-simplicial building blocks of geometry. The action functional commands their gluing such that $\rk$-dimensional discrete geometries can be generated by the quantum dynamics at a perturbative level. While the investigation of such theories' phase structure, has attracted considerable attention in recent years~\cite{Eichhorn:2013isa,Eichhorn:2014xaa,Eichhorn:2017xhy,Eichhorn:2018phj,Eichhorn:2018ylk,Eichhorn:2019hsa,Castro:2020dzt,Eichhorn:2020sla,Carrozza:2013oiy,Carrozza:2014rba,Benedetti:2015et,BenGeloun:2015ej,BenGeloun:2016kw,Benedetti:2016db,Carrozza:2016vsq,Carrozza:2016tih,Carrozza:2017vkz,BenGeloun:2018ekd,Pithis:2018bw,Baloitcha:2020idd,Pithis:2020sxm,Pithis:2020kio}, much remains to be understood concerning the general conditions under which critical behaviour occurs, and even more about whether this leads to well-defined continuum spacetime geometries.

The examination of the phase diagram of several GFT models in terms of the so-called functional renormalization group (FRG)~\cite{Berges:2002ga,Kopietz:2010zz,Dupuis:2020fhh} provides evidence for the formation of condensate phases in terms of (non-Gaussian) fixed points of the renormalization group flow (mostly in simplified models), and these have been given a tentative but promising continuum-geometric interpretation (for group field theories) in cosmological terms~\cite{Gielen:2016dss,Oriti:2016acw,Pithis:2019tvp}. However, similar to local theories, a phase of broken global symmetry, i.e. a condensate phase, can only be generated when the domain of the group field is non-compact (or its volume is sent to infinity in a thermodynamic limit), since otherwise the symmetry will be restored upon flow towards the IR~\cite{Pithis:2020sxm,Pithis:2020kio}. This had already been anticipated from corresponding mean-field analyses~\cite{Pithis:2018bw} and happens due to large fluctuations essentially generated by $\rk$-fold zero-modes in the spectrum on compact domain, see also~\cite{Benedetti1403}. 

Without such limiting procedure, i.e. maintaining that the non-local geometric degrees of freedom live on compact and finite domain, another way to tame the zero-mode effect could be to extend the field domain by non-compact directions. If one adds flat directions of local nature and interprets them as coordinates on an underlying flat space(time), one obtains the class of tensorial (group) field theory models used to reproduce the large-$N$ properties of SYK-type systems and possibly define new interesting conformal field theories~\cite{Rosenhaus:2018dtp,Delporte:2018iyf,Gurau:2019qag,Benedetti:2019ikb,Benedetti:2019rja,Benedetti:2020seh}. By doing so, however, one looses the (quantum) geometric interpretation of the models and the connection to quantum gravity is only (if these models define suitable CFTs) through the AdS/CFT correspondence. For recent work on phase transitions in such models, see for instance~\cite{Kim:2019upg,Klebanov:2020kck,Benedetti:2021qyk}.

In fact, coupling (discretized) matter to the group-theoretic quantum geometric models naturally leads to the same kind of extension of the field domain. For instance, coupling (free, massless) scalar fields to the geometric degrees of freedom results in the addition of $\mathbb{R}$-valued local variables to the domain of the group field~\cite{Oriti:2016qtz,Li:2017uao,Gielen:2018fqv}. Besides making the models more realistic, matter (in particular, scalar) fields can be naturally used to set up a material reference frame, allowing for a simple relational description (at least at an effective, continuum level~\cite{Oriti:2016qtz,Gielen:2018fqv,Marchetti:2020umh}) of the evolution of geometric quantities. 

For these reasons, it is pressing to understand in more detail the impact of matter degrees of freedom on the critical properties of such interacting matter-geometry systems, an issue that has so far been left aside in studies of the phase structure of GFTs (and indeed of general TGFTs). 

The general expectation, as anticipated, is that the addition of these non-compact local directions allows for a phase of broken symmetry which can be sustained throughout all scales, even if the discrete geometric degrees of freedom are encoded in a compact group domain. 
However, it is important to verify this expectation by a careful scrutiny of the phase properties of such TGFT models with mixed local/non-local degrees of freedom. This is the main objective of this work. We aim to do so by applying Landau-Ginzburg mean-field theory which is sufficient to point to the formation of a condensate phase. We focus on simplified TGFT models, here, as a first step toward the analysis of more realistic models, of direct relevance in particular for the GFT condensate cosmology program~\cite{Gielen:2016dss,Oriti:2016acw,Pithis:2019tvp}. 

\

To set the stage, let us briefly summarize the Landau-Ginzburg mean-field method applied to local scalar field theory~\cite{sachs_sen_sexton_2006,zinn2007phase}, which played a pivotal role for the Wilsonian renormalization group analysis of systems experiencing a continuous transition between a broken and an unbroken phase of a global symmetry~\cite{Wilson:1993dy}. One starts with the free energy functional of the system, written as an expansion in terms of (even) powers of a local field, the order parameter, and its gradient. One then considers a truncation thereof, which is assumed to be valid from the mesoscale to the macroscale, usually corresponding in terms of form to the (Euclideanized) classical action. Crucially, details about the microphysics are encoded in its coupling parameters and the order parameter. The latter is a coarse-grained quantity that features only universal properties of the system (e.g. the dimensionality of the underlying space and order parameter symmetries). This setting allows to control the thermodynamic phases of a system and to describe continuous phase transitions by studying long-range correlations of the order parameter fluctuations over a length $\xi$, the correlation length, beyond which correlations decay exponentially and which diverges at criticality. 

Concretely, in a first step one determines the (uniform) field configurations which minimize the free energy functional. In a second step, one studies correlations of quadratic fluctuations around this uniform background. This is the \textit{Gaussian approximation}~\cite{zinn2007phase}. To this aim, one linearizes the classical equations of motion using the fluctuations over the uniform background and then solves for the correlation function, in turn allowing to compute the correlation length. Finally, the domain of validity of this approximation scheme can be studied by quantifying the strength of the fluctuations. As long as these remain small and the interaction term can be treated perturbatively, mean-field theory around the Gaussian theory can be self-consistently applied. This is the \textit{Ginzburg criterion}~\cite{Kopietz:2010zz}. In particular, it allows to establish the well-known result that the critical dimension below which mean-field theory ceases to provide an accurate description of the phase diagram and of the (second-order) phase transition is $4$. Right below that value, mean-field theory still captures the basic structure of the phase diagram, but only qualitatively, and the numerical results on the critical exponents are misleading in general~\cite{zinn2007phase}.\footnote{In mean-field theory, only fluctuations on the scale $\xi$ are considered important in the vicinity of the phase transition. In order to account for the correct quantitative critical behavior, fluctuations on all scales have to be considered at the phase transition, and thus one has to resort to the full renormalization group machinery~\cite{Wilson:1993dy}.}  

In this work, we adapt and apply the Landau-Ginzburg method to TGFTs that include also local (from the TGFT point of view, not necessarily with respect to spacetime physics) degrees of freedom. The main challenge (as for TGFT renormalization) is to deal with  the non-locality of the geometric degrees of freedom and with their mixing with the local ones.

\

The article is organized as follows. In Section \ref{sec:LGT} we introduce tensorial group field theory, including additional local non-compact degrees of freedom, clarify the relevant theory space and detail Landau-Ginzburg mean-field theory in this context. This allows us to determine the correlation function of fluctuations over uniform background configurations. In Section \ref{sec:correlationlength} we use this result to compute the correlation length considering both compact and non-compact limits of the non-local geometric degrees of freedom. In Section \ref{sec:ginzburg} we establish, via the Ginzburg criterion, the conditions under which mean-field theory is valid, i.e. we determine the critical dimension for arbitrary tensor rank $\rk$, compact vs. non-compact Abelian Lie group $G$, $\std$ local degrees of freedom and interactions compatible with the global symmetry to be broken via second-order phase transition. In the final Section \ref{sec:conclusion} we give a synopsis of our results, discuss limitations of our analysis and give an outlook for future research. Appendix \ref{app:integrals} complements the main Sections with explicit computations of the correlation functions. 

\section{Landau-Ginzburg theory for TGFTs with local directions}\label{sec:LGT}

In the following Section, we set up Landau-Ginzburg mean-field theory in the context of TGFTs also including local directions as used to define relational frames. Effectively, these are hybrid theories with local and non-local degrees of freedom which requires a modified regularization scheme and a careful discussion of the notion of correlation length, before the critical dimension can be deduced via the Ginzburg criterion thereafter.  

\subsection{Tensorial group field theory with local directions}

Tensorial group field theory, before the introduction of locally coupled \lq matter\rq~degrees of freedom, is characterized by a configuration space $\gm^{\times\rk}\equiv \gm^\rk$ of $\rk$ copies of a Lie group~$\gm$ and combinatorially non-local interactions.
Frame coordinates $\ff\in\R^\std$ extend the configuration space to $\R^\std \times \gm^{\rk}$ but the interactions are local in these variables\footnote{Since TGFT models generalize matrix models for $2d$ gravity~\cite{DiFrancesco:1993cyw} to higher dimensions, rank-$r$ group fields correspond to $r-1$-simplices (representing the fundamental building blocks of geometry) and the interactions encode how to glue these together to build $r$-dimensional discrete geometries. Introducing at the level of classical discretized gravity a discretized scalar field associates a scalar field value to each $r-1$-simplex. Since the scalar field values are then identified at the level of a 4-simplex dual to a TGFT Feynman diagram, the interaction kernel is then characterized by a single value of the scalar field, which is what we mean here by locality. See~\cite{Oriti:2016qtz,Li:2017uao,Gielen:2018fqv} for a detailed discussion.}. Such fields are often introduced in the background-independent contexts of classical and quantum gravity~\cite{DeWitt:1962cg,Brown:1994py,Rovelli:2001bz,Dittrich:2005kc,Giddings:2005id,Giesel:2012rb,Li:2017uao,Gielen:2018fqv}, in order to be used as physical reference frames allowing to describe the relational evolution of physical quantities preserving diffeomorphism invariance. This is also how they have been used in GFT cosmology. In light of this common usage, we label the new local directions \lq frame coordinates\rq.

Fields $\gf:\R^\std \times\gm^{\rk}\to\mathbb{R}$ or $\mathbb{C}$ therefore depend on 
$\std$ arguments $\vff=(\phi_1,...,\phi_\std)$ each in $\R$ and 
$\rk$ arguments $\vg = (g_1 , g_2 , ..., g_\rk)$ each in a Lie group $G$.\footnote{Hereafter, we focus on real-valued fields noting that the main results in the remainder of this article are easily transferred to complex-valued TGFT fields.}
They are square-integrable functions $\gf,\gf' \in L^2(\R^\std \times G^{\rk})$ with respect to the inner product
\begin{equation}
(\gf,\gf') = \int_{\R^\std} \d\vff \int_{\gm^\rk} \d\vg\, \gfb(\vff,\vg) \gf'(\vff,\vg)\,,
\end{equation}
defined in terms of the Haar measure, which we take to be unnormalized. In this Section, we will restrict to compact Lie groups only, because, as we will see below, this allows to carry on with the Landau-Ginzburg analysis with a homogeneous mean-field ansatz without encountering unphysical divergences (we will take care of this issue, instead, in the following sections). When $\gm$ is compact, the group volume
\begin{equation}
\int_\gm \d g = \cs \,
\end{equation}
is finite and may be seen to represent an IR regularization from the formal field-theoretic point of view (i.e. no spacetime interpretation is assumed)~\cite{BenGeloun:2015ej,BenGeloun:2016kw}. In this sense, taking a large-volume limit $\cs\rightarrow\infty$ allows us to explore the non-compact limit of the model.\footnote{This can be compared with the use of the Wick rotation in ordinary QFTs on $\mathbb{R}\times \mathbb{R}^3$ together with the imposition of periodic (or anti-periodic) boundary conditions (motivated by the assumption that field configurations approach an asymptotic form at $t\to \pm\infty$ and can thus be identified). Such theories are thermal field theories on $S^1\times \mathbb{R}^3$, where the radius of the compactified dimension is proportional to the inverse temperature $\beta$. Taking the zero-temperature limit, that is, by sending $\beta\to\infty$, it is commonly understood that the compactification of $\mathbb{R}$ to $S^1$ is undone, so that one is led back to the standard QFT formulation~\cite{Zinn-Justin:2000ecv,Zinn-Justin:2002ecy}. In our context, the large-volume limit thus allows us to relate the theory on $\gm=\text{U}(1)$ to the one on $\gm=\mathbb{R}$.}

For such fields, the \enquote{Fourier} transform in group space is the expansion in the matrix coefficients of unitary irreducible representations labelled by a multi-index $\vrep=(\rep_1,...,\rep_\rk)$
\begin{equation}\label{eq:transform}
\gf(\vff,\vg) = \sum_{\rep_1,...,\rep_\rk} \left(\prod_{\ell=1}^\rk \frac{d_{\rep_\ell}}{\cs}\right) \tr_\vrep \left[ \gf(\pmb{\phi},\pmb{j})\bigotimes_{\ell =1}^\rk D^{\rep_\ell}(g_\ell) \right],
\end{equation}
where $D^{\rep_\ell}(g_\ell)$ are the representation matrices on $d_{j_\ell}$-dimensional representation space. Their coefficients form a countable complete orthogonal basis of $L^2(G)$ according to the Peter-Weyl theorem. 
Thus, also the field transform $\gf(\pmb{\phi},\pmb{j})$ is matrix-valued with respect to each representation $\rep_\ell $ and $\tr_\vrep$ is the trace over all the representation spaces~\cite{jacques2008analysis}. Finally, standard Fourier transform on $\mathbb{R}^\std$ relates the reference frame representation of the field $\gf(\pmb{\phi},\pmb{j})$ to that on its conjugate \lq momentum\rq~space $\vfm$, i.e.,
\begin{equation}
    \gf(\pmb{\phi},\pmb{j})=\int_{\mathbb{R}^\std}\frac{\text{d}^\std \fm}{(2\pi)^{\std}}\gf(\vfm,\pmb{j})\text{e}^{i\vff\cdot \vfm},
\end{equation}
where $\vff\cdot \vfm\equiv \sum_{i=1}^\std \fm_{i}\phi_i$.\footnote{Notice that we are assuming a Euclidean signature for the frame coordinates. Although non-Euclidean signatures may be important for cosmological applications of these models (see e.g.~\cite{Oriti:2016qtz,Marchetti:2020umh}), using a Euclidean signature is customary for applications of mean-field theory to statistical systems, and guarantees that the uniform mean-field configuration, which we will employ below, indeed minimizes the action. Thus we stick to this simplifying choice.}

\

The combinatorial non-locality of TGFT interactions is encoded in the way field arguments on $G$ are paired. That is, while the interaction of the frame coordinates $
\vff$  has the usual local, i.e. point-like, form (with a single integration $\d\vff$), the interaction of the non-local group degrees of freedom at a vertex is a convolution encoded by a graph $\gamma$, 
\begin{equation}\label{eq:nonlocalinteraction}
\int_{\R^\std}\! \d \vff\, 
\Tr_\gamma(\gf) := \int_{\R^\std}\! \d\vff
\int_{\gm^{\times\rk\cdot\nv_\gamma}} \prod_{i=1}^{\nv_\gamma} \d\vg_i\,  \prod_{(i,a;j,b)}\delta(g_i^a/g_j^{b}) \prod_{i=1}^{\nv_\gamma} {\gf}(\vff,\vg_i)\,,
\end{equation}
where $\nv_\gamma$ is the number of vertices of $\gamma$ and the product of Dirac distributions\footnote{Note that, by choosing simple delta distribution kernels for the interaction, we are choosing a simple class of TGFT models, simpler, for example, than the quantum geometric models encoding a richer geometry at the discrete level.} is over its edges labelled $(i,a;j,b)$ where $i,j=1,2,...,\nv_\gamma$ and $a,b=1,2,...,
\rk$.  For example, for the melonic graph $\gamma =\vertexexvg$ (labelling the vertices $i,j=1,2,3,4$ and the group arguments $a,b=1,2,3$) the interaction term in the action would be

\begin{equation}\label{eq:interactionexample}
\int_{\R^\std}\! \d \vff\, 
\Tr_{\vertexexvg}(\gf) = 
\int_{\R^\std}\! \d\vff
\int_{\gm^{\times 3\cdot 4}} \prod_{i=1}^{4} \d\vg_i\!  \prod_{\substack{(ij)=\\(14),(23)}}\!\delta(g_i^1/g_j^1)
\prod_{\substack{(ij)=\\(12),(34)}}\!\delta(g_i^2/g_j^2)\delta(g_i^3/g_j^3)
\prod_{i=1}^{4} {\gf}(\vff,\vg_i) .
\end{equation}
There are various ways to represent such an interaction vertex diagrammatically, for example as a stranded vertex or directly using the vertex graph~\cite{Oriti:2015kv,Thurigen:2021vy},
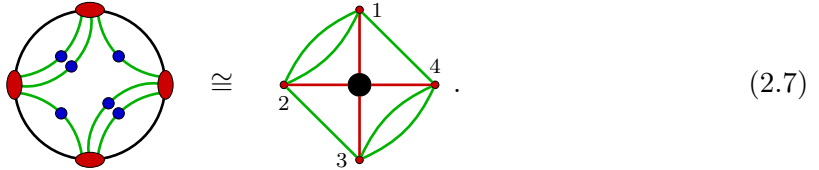
\begin{equation}
\vertexexstranded \quad \cong \quad \vertexexgraph \, .
\end{equation}
Here, red vertices labelled $i=1,2,3,4$ represent the fields $\gf(\vff,\vg_i)$ and the pairing of green half edges corresponding to the single arguments $g_i^a$ into edges encodes their pairwise convolution.\footnote{As usual, in momentum space the point-like interaction with respect to $\vff$ results in a convolution $\delta(\sum_{i=1}^{\nv_\gamma}k_i)$ (black vertex) of momenta $k_i$ (red half-edges). 
Anyhow, this vertex and adjacency to half edges provides necessary information even for purely non-local interactions as it encodes which connected components belong one multi-trace vertex, e.g. $\gamma=\gamma_1\sqcup\gamma_2$, since $\Tr_{\gamma_1\sqcup\gamma_2} = \Tr_{\gamma_1}\cdot\Tr_{\gamma_2}$ factorizes but is still one interaction vertex (see~\cite{Thurigen:2021vy} for further details).
}

The TGFT action on $\gm^\rk$ with $\std$ frame coordinates has thus the general form
\begin{equation}\label{eq:action}
S[\gf] = \left(\gf, \kin\gf\right) + \sum_\gamma \lambda_\gamma \int_{\R^\std}\!\d\vff\, \Tr_\gamma(\gf)\,,
\end{equation}
where the sum runs over a set of vertex graphs $\gamma$ which specify the theory together with a particular kinetic term $\kin$.
In the following, we will consider
\begin{equation}\label{eq:mixedplaplacian}
\kin = -\sum_{i=1}^\std \alpha_i
\partial^2_{\ff_i} + \sum_{\ell=1}^\rk (-1)^\gd \Delta_\ell + \mu \,,
\end{equation}
where $\Delta_\ell$ is the Laplace operator on the $\gd$-dimensional Lie group $\gm$ and $\alpha_i$ is assumed to be a positive function of the group variables, reflecting the non-trivial details of the coupling of the scalar fields to the geometry~\cite{Oriti:2016qtz,Li:2017uao}.\footnote{In general, such coupling results in a form of the kinetic kernel which is non-local with respect to each $\ff_i$ variable. More precisely, for each $\ff_i$, the kinetic kernel can be explicitly written in terms of a series of second derivatives with respect to $\ff_i$ acting on a delta-function of the same argument \cite{Li:2017uao}. In the spirit of Landau-Ginzburg theory, however, in \eqref{eq:mixedplaplacian} we only include terms of the aforementioned series with the lowest number of derivatives. Such a derivative truncation has been for instance shown to be justified in models where the matter degrees of freedom are used as material reference frames~\cite{Marchetti:2020umh}.} Notice that in~\eqref{eq:mixedplaplacian} the coupling is encoded in the $\alpha_i$, like refractive indices for light propagation in anisotropic media. We will assume from now on that different frame coordinate directions are weighted equally, meaning $\alpha_i\equiv \cc$ for each $i=1,\dots, \std$. Finally, we note that~\eqref{eq:action} is endowed with a global $\mathbb{Z}_2$-symmetry. Landau-Ginzburg theory seeks to characterize a continuous transition between a broken (i.e. condensate) and unbroken phase of this symmetry.

\subsection{Gaussian approximation}\label{subsec:gaussianapprox}

For the Ginzburg criterion one has to determine the $2$-point correlation function in the Gaussian approximation. To this end, one considers correlations of fluctuations $\pgf$ over a constant background $\cgf$, also referred to as the mean order parameter.

To start with, one derives the equations of motion from the general action~\eqref{eq:action} giving
\begin{equation}\label{eq:eom}
\kin \gf + \sum_\gamma \lambda_\gamma \sum_{v\in\V_\gamma} \Tr_{\gamma\setminus v}(\gf) = 0\,,
\end{equation}
where the last sum runs over all vertices in the graph's vertex set $\V_\gamma$ of traces encoded by the graph $\gamma\setminus v$ which is obtained by deleting the vertex $v$ (together with its adjacent stranded half edges).  
Thus, in this convolution there is one field less.
Take again the interaction $\gamma =\vertexexvg$,~\eqref{eq:interactionexample}, as an example. 
Its four vertices are completely symmetric, such that 
\begin{equation}
\lambda_{\vertexexvg} \sum_{v\in\V_\gamma} \Tr_{\gamma\setminus v}(\gf)
= 4 \, \vertexexeom{\gf(\vff,\vg_1)}{\gf(\vff,\vg_2)}{\gf(\vff,\vg_3)}
= 4 \lambda_{\vertexexvg}
\int_{\gm^{6}} \prod_{i=1}^{2} \d\vg_i\,  \delta(g_2^1/g_3^1)
\delta(g_1^2/g_2^2)\delta(g_1^3/g_2^3)
\prod_{i=1}^{3} {\gf}(\vff,\vg_i) 
 \, .
\end{equation}
 Projecting to the constant field $\gf(\vff,\vg)=\cgf$ leaves as many empty group integrals and thus volume factors $\cs$, as there are internal edges in $\gamma\setminus v$. The number $\nei_{\gamma\setminus v}$ of these edges relates to the number of vertices as $2\nei_{\gamma\setminus v} = \rk(\nv_{\gamma\setminus v}-1)=\rk(\nv_\gamma-2)$ such that the equation for $\cgf$ is
\begin{equation}\label{eq:eomconstant}
\mu\cgf + \sum_\gamma \lambda_\gamma \nv_\gamma \cs^{\rk\frac{\nv_\gamma-2}{2}} \cgf^{\nv_\gamma-1}
= \left(\mu+\sum_\gamma \lambda_\gamma \nv_\gamma \cs^{\rk\frac{\nv_\gamma-2}{2}} \cgf^{\nv_\gamma-2} \right)\cgf = 0\,.
\end{equation}
The solution $\cgf=0$ factors trivially.
The remaining part is in general an algebraic equation of the order of two less than the highest-order interaction\footnote{Notice that if $\gm$ was non-compact the factor $\cs^{\rk(\nv_\gamma-2)/2}$ would be diverging. This factor appears because the homogenous mean-field ansatz together with the non-local nature of interactions which produces exactly $\rk(\nv_\gamma-2)/2$ empty integrals.}. 
In particular, for a sum of interactions each given by vertex graph $\gamma$
with the same number of vertices $\nv_\gamma=\ooi$
the solutions simply are the $i=1,2,...,\ooi_\gamma-2$ roots 
\begin{equation}\label{eq:nonzerosolution}
\cs^{\frac{\rk}{2}}\cgf = \zeta_i \left(-\frac{\mu}{\ooi \sum_\gamma \lambda_\gamma} \right)^{\frac{1}{\ooi-2}}\,,
\end{equation}
where $\zeta_i$ is the $i$'th root of unity.
Note that the presence of the volume factor in front of~$\cgf$ is completely natural since the relevant argument of the interaction potential in the full renormalization group flow is~\cite{Pithis:2020kio,Pithis:2020sxm}
\begin{equation}
\rho := \int_{G^\rk}\cgf^2 = \cs^\rk \cgf^2 \, .
\end{equation}
For a sum over quartic interactions, $\ooi=4$, this solution is simply

\begin{equation}\label{eq:quarticvacuum}
\rho = {-\frac{\mu}{4\sum_\gamma\lambda_\gamma}} \, .
\end{equation}
The standard Landau-Ginzburg analysis associates the solutions
\begin{equation}\label{eq:minimum}
\cgf=0 \textrm{ to } \mu>0 \quad\textrm{ and } \quad
\cs^{\frac{\rk}{2}}\cgf = \pm \sqrt{-\frac{\mu}{4\sum_\gamma\lambda_\gamma}} \textrm{ to } \mu<0 \,.
\end{equation}
For the analysis of phase transitions one is interested in the \enquote{broken} phase $\mu<0$.
This is the phase for which the global $\mathbb{Z}_2$-symmetry is broken and the mean-field order parameter $\cgf$ is non-vanishing, corresponding to a non-zero local minimum of the interaction potential.

\ 

In the so-called Gaussian (or quasi-Gaussian \cite{zinn2007phase}) approximation one considers small fluctuations $\pgf$ around the uniform background $\cgf$, that is 
\begin{equation}
\gf(\vff,\vg)=\cgf+\pgf(\vff,\vg) \, .
\end{equation}
With this ansatz, the first order (in $\delta\Phi$) equations of motion are
\begin{equation}\label{eq:eomGaussian}
\kin \pgf + \sum_\gamma \lambda_\gamma \sum_{v,v'\in\V_\gamma} \Tr_{\gamma\setminus v}(\cgf,\pgf_{v'}) = 0\,,
\end{equation}
where the second argument $\pgf_{v'}$ in the trace now means that the field $\pgf$ is inserted at the vertex $v'\in \gamma\setminus v$ (and $\cgf$ at all other vertices).
In the example of~\eqref{eq:interactionexample}, there is now a non-trivial sum over the remaining three vertices $v'$,
\begin{align}
&\frac{1}{4}\lambda_{\vertexexvg} \sum_{v,v'\in\V_\gamma} \Tr_{\gamma\setminus v}(\cgf,\pgf_{v'})
=\vertexexeom{\pgf(\vff,\vg_1)}{\cgf}{\cgf} 
\,+\,\vertexexeom{\cgf}{\pgf(\vff,\vg_2)}{\cgf}
\,+\,\vertexexeom{\cgf}{\cgf}{\pgf(\vff,\vg_3)} \\
&= \lambda_{\vertexexvg} \cgf^2 \left(
\cs \int_{\gm^2} \d g_1^2\d g_1^3 \,\pgf(\vff,\vg_1)
+ \int_{\gm^3} \d g_2^1\d g_2^2\d g_1^3 \,\pgf(\vff,\vg_2)
+ \cs^2\int_{\gm} \d g_3^1 \,\pgf(\vff,\vg_3) \nonumber
\right) \, .
\end{align}
Relabelling the free group arguments simply as $(g_1^1,g_3^2,g_3^3)=(g^1,g^2,g^3)=:\vg$ and writing the expression as a quadratic form acting on a single $\pgf$ in terms of a single convolution kernel, we can write
\begin{equation}\label{eq:exampleHessian}
\sum_{v,v'\in\V_{\vertexexvg}}\! \Tr_{\vertexexvg\setminus v}(\cgf,\pgf_{v'})
=4\cgf^2 \int_{G^3} \!\!\! \d\vh \bigg\{
\cs \delta(g^1/h^1) + 1 + \cs\delta(g^2/h^2)\cdot\cs\delta(g^3/h^3)
\bigg\} \pgf(\vff,\vh) \, .
\end{equation}
This quadratic form is simply the Hessian of the interaction part of the action, occurring also in the Wetterich-Morris functional renormalization group equation~\cite{Berges:2002ga,Dupuis:2020fhh,Kopietz:2010zz}, with the only difference that here it is simply taken from the action $S$ while there it is defined from the so-called effective average action $\Gamma_k$.
Still, both are of the same form and differ only in that here the couplings $\lambda_\gamma$ are fixed, while there they are dynamically depending on the renormalization group scale $k$.
In particular, the example~\eqref{eq:exampleHessian} has already been calculated in~\cite{Pithis:2020kio,Pithis:2020sxm} as it is the particular quartic case of a so-called cyclic-melonic interaction (a closed chain of open melons). The structure is much more general though.

Thus, the general equations of motions for the field perturbations $\pgf$ in the Gaussian approximation~\eqref{eq:eomGaussian} are
\begin{equation}\label{eq:eomGaussian2}
(\kin + F[\cgf])\pgf(\vff,\vg) = 0\, .
\end{equation}
Therein, the Hessian of the interaction part has the general form 
\begin{equation}
F[\gf](\vff,\vg;\vff',\vh):=\frac{\delta S_\textsc{ia}[\gf]}{\delta\gf(\vff,\vg)\delta\gf(\vff',\vh)}
= \delta(\vff-\vff')\sum_\gamma \lambda_\gamma \sum_{v,v'\in\V_\gamma} \Tr_{\gamma\setminus v\setminus v'}(\gf)
\end{equation}
and is evaluated at $\gf(\vff,\vg)=\cgf$ in the perturbed equation of motion (\ref{eq:eomGaussian2}). As a consequence, there is a Dirac delta distribution in the trace $\Tr_{\gamma\setminus v\setminus v'}(\cgf)$ for each edge connecting the vertices $v$ and $v'$.
In general, this leads to terms with various combinations of Dirac distributions as in the example

\begin{equation}
F[\cgf](\vff,\vg;\vff',\vh) = 4\lambda_{\vertexexvg} \cgf^2 \delta(\vff-\vff') \bigg\{
1 + \cs \delta(g^1/h^1) + \cs\delta(g^2/h^2)\cdot\cs\delta(g^3/h^3)
\bigg\} \, .
\end{equation}
In more symmetric cases, only particular products of delta distributions occur.  For example, for the simplicial interaction given by the complete graph with $\rk$ vertices $\gamma = K_\rk$ one has
\begin{equation}
F[\cgf](\vff,\vg;\vff',\vh) = \rk\lambda_{K_\rk} \cgf^{\rk-2} \delta(\vff-\vff') \sum_{\ell=1}^\rk \cs\delta(g^\ell/h^\ell) \, ,
\end{equation}
as investigated for $\rk=3$ (the Boulatov model~\cite{Boulatov:1992vp}) in~\cite{Pithis:2018bw}.
In general, for a sum over interactions of the same order $\ooi$, that is each  with vertex graph $\gamma$ with $\nv_\gamma=\ooi$,  one can insert the non-vanishing solution~\eqref{eq:nonzerosolution} to find
\begin{align}\label{eq:projectedinteractiongroupspace}
F[\cgf](\vff,\vg;\vff',\vh) &=  \cs^{\rk\left(\frac{\ooi}{2}-2\right)}\cgf^{\ooi-2} \delta(\vff-\vff') \sum_\gamma \lambda_\gamma \nlop_\gamma(\vg,\vh) 
= - \mu\,  \delta(\vff-\vff') \frac{1}{\cs^{\rk}} \sum_\gamma \ct_\gamma\nlop_\gamma(\vg,\vh),
\end{align}
where $\nlop_\gamma(\vg,\vh)$ is a sum over products of Dirac distributions specific to the combinatorial structure of $\gamma$ (see Tab.~\ref{tab:vertexexamples} for further examples)
and
\[\label{eq:averagedcoupling}
\ct_\gamma = \frac{\lambda_\gamma}{\ooi \sum_{\gamma'}\lambda_{\gamma'}}
\,.\] 
Note that $\cs^{-\rk}$ is simply the natural explicit volume factor for a quadratic form.

\begin{table}[]
    \centering
    \begin{tabular}{c|c|l}
         double-trace melon & \cvft & $\nloph_{\cvtvg\bigsqcup\cvtvg}(\vrep)=4\left(2\prod\limits_{\ell=1}^4\delta_{\rep_\ell,0}+1\right)$
        \\ \hline
        quartic melonic    & \cvf  & 
        $\nloph_{\cvfvg}(\vrep)= 4\left(\prod\limits_{\ell}\delta_{\rep_\ell,0}+\prod\limits_{b\ne\ell}\delta_{\rep_b,0}+\delta_{\rep_\ell,0}\right)$ 
        \\ \hline
        quartic necklace & \cvfn & 
        $\nloph_{\cvfnvg}(\vrep)= 4\left(\prod\limits_{\ell}\delta_{\rep_\ell,0}+ \delta_{\rep_1,0}\delta_{\rep_2,0} +\delta_{\rep_3,0}\delta_{\rep_4,0}\right)$
        \\ \hline
        simplicial
        & \cvfs &
        $\nloph_{\cvfsvg}(\vrep)= 5 \sum\limits_{i=0}^4 \prod\limits_{k\ne i} \delta_{\rep_{(ik)},0}$ \scriptsize{ (edges labeled by adjacent vertices $i,k$)}
    \end{tabular} 
    \caption{Examples of non-local interaction vertex graphs for $\rk=4$ group arguments and the resulting operator $\nloph_\gamma$ in representation space.}
    \label{tab:vertexexamples}
\end{table}

As a general result, non-locality leads to a quadratic (\qmark{Gaussian}) term $\kin+F[\cgf]$ in which not only the second derivative in the kinetic term $\kin$ but also the remainder of the interactions $F[\cgf]$ are not diagonal in group space.
To have a diagonal form, $F[\cgf]$ would need to be proportional to $\prod_\ell \delta(g^\ell/h^\ell)$.
This is already known from the FRG analysis~\cite{Pithis:2020kio,Pithis:2020sxm}.
The standard way to diagonalize $\kin$ is to transform to momentum (representation) space. For the Dirac distribution, this transformation gives simply one, that is
\begin{equation}
\delta(g) = \sum_j \frac{d_j}{\cs} \tr_j D^j(g) \,.
\end{equation}
Thus, each term with a Dirac delta distribution for $p$ of the $\rk$ group arguments gives $\rk-p$ zero-modes in representation space.

In this way one obtains the representation-space correlation function in the Gaussian approximation. For the sum over interactions of the same order~$\ooi$, \eqref{eq:projectedinteractiongroupspace}, one has
\begin{equation}\label{eq:projectedinteractionrepspace}
\hat{F}[\cgf](\vfm,\vrep;\vfm',\vrep')
= -{\mu}\delta(\vfm + \vfm'){\sum_\gamma}\ct_\gamma \nloph_\gamma(\vrep)    \prod_{\ell=1}^\rk \delta_{\rep_\ell,\rep'_\ell} \one_{\rep_\ell}
\,,
\end{equation}
where $\one_{\rep}$ is the unit matrix in the $\rep$ representation space and

\begin{equation}\label{eq:nonlocaloperator}
\nloph_\gamma(\vrep) = \sum_{p=0}^\rk \sum_{(\ell_0,...,\ell_p)} \nlc_{\ell_0...\ell_p} \prod_{\ell=\ell_1}^{\ell_p} \delta_{\rep_{\ell,0}}
\end{equation}
with combinatorial factors $\nlc_{\ell_0...\ell_p}$ specific to the combinatorial structure of each vertex graph $\gamma$ (see Tab.~\ref{tab:vertexexamples}). 
A contribution $\nlc_{\ell_0...\ell_\rk}\ne0$ occurs for any interaction except for simplicial ones, that is those given by a complete graph $\gamma=K_\rk$.

Since the kinetic operator in representation space is also diagonal,
\begin{equation}
\hat{\kin}(\vfm,\vrep;\vfm',\vrep') = \left(\cc(\vrep)\sum_{i=1}^\std \fm_i^2 + \frac1{\cs^2}\sum_{\ell=1}^\rk \cas_{\rep_\ell} +\mu\right)
\delta(\vfm+\vfm')\prod_{\ell=1}^\rk \delta_{\rep_\ell,\rep'_\ell} \one_{\rep_\ell}\, ,
\end{equation}
one may easily obtain the $2$-point correlation function in the Gaussian approximation as
\begin{equation}\label{Fouriercoefficients}
\hat{C}(\vfm,\vrep) = (\hat{\kin}+\hat{F}[\cgf])^{-1}(\vfm,\vrep) 
= \frac{\one_{\rep_\ell}}{\cc(\vrep)\sum_a
\fm_a^2+\frac1{\cs^2}\sum_\ell\cas_{\rep_\ell} + \mu - {\mu}\sum_\gamma \ct_\gamma\nloph_\gamma(\vrep)} \, .
\end{equation}
It is the specific property of non-local interactions that the effective mass
\begin{equation}\label{eq:effectivemass}
b_\vrep := \mu\bigg(1 - \sum_\gamma \ct_\gamma\nloph_\gamma(\vrep)\bigg)
\end{equation}
is not constant, in contrast to usual local field theories, but depends on the combinatorics of the non-local interactions, see also Tab.~\ref{tab:vertexexamples}.

\paragraph{Considering the closure constraint.}

In the quantum geometric GFT models one often imposes a specific symmetry onto the fields, the so-called closure or gauge constraint, that is 
\begin{equation}\label{eq:fieldgaugeconstraint}
\Phi(\pmb{\phi},g_1,...,g_r)=\Phi(\pmb{\phi},g_1 h,...,g_r h)~~~\forall h\in G
\end{equation}
which is typically imposed via group averaging. It corresponds to a physical enrichment of the models in that it is part of a more complete geometric characterization of the TGFT quanta. In particular, it implies the closure of the flux variables which are dual to the group elements in~\eqref{eq:fieldgaugeconstraint}, a necessary condition for a geometric $r-1$-simplex. It can be motivated also in relations to loop gravity~\cite{Ashtekar:2004eh,Rovelli:2011tk}, see also~\cite{Baratin:2010nn,Guedes:2013vi}, and it implies as well that the Feynman amplitudes of the TGFT model takes the form of a lattice gauge theory partition function, on the lattice dual to the Feynman diagram \cite{Carrozza:2016tih}. The \enquote{Fourier} transform in group space~\eqref{eq:transform} of the field is thus modified to
\begin{equation}\label{eq:transformgc}
\gf(\vff,\vg) = \sum_{\rep_1,...,\rep_\rk} \left(\prod_{\ell=1}^\rk \frac{d_{\rep_\ell}}{\cs}\right) \tr_\vrep \left[ \gf(\pmb{\phi},\pmb{j})\int\text{d}h\bigotimes_{\ell =1}^\rk D^{\rep_\ell}(g_\ell h) \right]
\end{equation}
and the $2$-point function in representation space simply yields 
\begin{equation}\label{Fouriercoefficientsgc}
\hat{C}(\vfm,\vrep) 
= \frac{\int\text{d}h\bigotimes_{\ell =1}^\rk D^{\rep_\ell}( h)}{\cc(\vrep)\sum_i
\fm_i^2+\frac1{\cs^2}\sum_\ell\cas_{\rep_\ell} + b_\vrep} \, .
\end{equation}
Since the ensuing computations go analogously through for configurations subject to the closure constraint, we carry on with the discussion for the general field configuration, and then comment towards the end on how the value for the critical dimension changes when this condition is imposed. In foresight, the naive expectation that the closure reduces the rank by one, in turn leading to one additional zero-mode, will prove to be correct.

\section{Correlations on Abelian group manifolds}\label{sec:correlationlength}

In condensed-matter statistical systems, the correlation length sets the scale beyond which correlations die off exponentially, thus providing a characteristic scale for fluctuations. Importantly, as we will discuss in Section \ref{sec:ginzburg}, it is instrumental for the evaluation of the so-called Ginzburg $Q$ factor which measures the strength of fluctuations in the context of Landau-Ginzburg mean-field theory and thus provides crucial information on the occurrence of phase transitions as well as the domain of validity of mean-field theory. 
The notion of correlation length for TGFTs was first discussed in Ref.~\cite{Pithis:2018bw}. Here we provide a definition of it for Abelian\footnote{This definition can be suitably generalized to more complicated group structures, as we mention in Section \ref{subsec:compactcase} and as it is discussed in more detail in~\cite{Marchetti:2021xyz}.} 
group manifolds $G$,
which is motivated by the analogous definition standard statistical systems. In Section \ref{subsec:compactcase}
we compute the correlation length on compact and Abelian group manifolds $G$
from this definition. 
In Section \ref{subsec:explicitexamplescorr}, we repeat the analysis considering instead the non-compact Abelian case, discussing also the features of the correlation function in \qmark{coordinate space}, which is directly relevant for the evaluation of the Ginzburg $Q$ factor thereafter in Section~\ref{sec:ginzburg}.

\subsection{Compact case}\label{subsec:compactcase}

Before discussing in detail the definition of correlation length, let us clarify the setting and the notation that we will use below. The field theories considered here live on the configuration space $D\cong\mathbb{R}^\std\times G^r$, where $G$ is an Abelian, compact and connected Lie group. It is a classical result~\cite{ref1} that any such $G$ is isomorphic to $\text{U}(1)^{d_G}$, with $d_G=\dim G$. So, from now on we will consider $G^r\cong \text{U}(1)^{\ed}$, with $\ed\equiv rd_G$. Furthermore, we parametrize $\text{U}(1)$ by means of the coordinate $\theta\in[-\pi \rvp,\pi \rvp]$, so that the volume of each $\text{U}(1)$ factor is $\rv\equiv 2\pi \rvp$ and the group volume is $\cs\equiv \rv^{d_G}\equiv (2\pi\rvp)^{d_G}$. Correspondingly, $G^r$ is parametrized by $\pmb{\theta}=\{\vec{\theta}_1,\dots,\vec{\theta}_{r}\}$, with $\vec{\theta}_\ell=\{\theta_{\ell,1},\dots,\theta_{\ell,d_G}\}$. When considering $\pmb{\theta}$ as an element of $\text{U}(1)^\ed$, the components of each $\vec{\theta}_\ell$ are mapped into the components $\{\theta_{\ell d_G+1},\dots, \theta_{(\ell+1)d_G}\}$ of $\pmb{\theta}=\{\theta_1,\dots,\theta_s\}$. We will switch from one notation to the other one when more convenient.

\paragraph{Definition of the correlation length.}
In statistical field theory, the (square of the) correlation length can be defined as the Taylor coefficient of the susceptibility at order two in the momenta. Equivalently, being the susceptibility just the Fourier transform of the correlation function, one can define the correlation length as the second moment of the correlation function~\cite{doi:10.1002/9783527603978.mst0387}. While this definition is arguably very natural, there are two main points to keep in mind, when adapting it to our context. 
First, field theories arising in statistical and condensed matter physics are usually local, while in the present context we are dealing with field theories with non-local interactions. 
Second, the domain of the field theories we are considering here has no direct spacetime interpretation, so any notion of correlation length cannot be associated to distances in physical space, but only understood as identifying some (important) internal scale. In what follows, we maintain the above definition for the correlation length as second moment of the correlation function, though we are aware that further scrutiny into the physics of our non-local and pre-geometric field theories might suggest modifications in the future\footnote{In this regard, we point out that we have made the choice of considering our TGFT models on group manifolds treating such domain as the direct analogue of configuration space in usual QFTs on spacetime, and the group modes as the analogue of momenta. This is natural, in accordance with the usual RG approach to TGFTs and with the appearance of differential operators on the group in their kinetic terms, and the easiest choice from a formal point of view. However, a priori one could also take the opposite perspective, consider the dual group algebra as the relevant configuration space, view TGFTs as QFTs on a non-commutative (for non-Abelian groups) manifolds and a curved momentum space, and proceed accordingly. Some work in the TGFT literature takes indeed this perspective, so also this possibility should be left for consideration for future work.}.

The first step to obtain a formula for the correlation length~\cite{doi:10.1002/9783527603978.mst0387} is to write down the Fourier transform of the field, i.e.
\begin{equation}\label{eqn:firststepderivation}
    \hat{\Phi}(\vfm,\pmb{n})
    =\int\text{d}^{\ed} \theta\text{d}^\std\phi\, \text{e}^{-i\pmb{\phi}\cdot\vfm}e^{-i\pmb{\theta}\cdot \pmb{n}/a}\Phi(\pmb{\phi},\pmb{\theta})\,,
\end{equation}
where $\pmb{\theta}\cdot \pmb{n}=\sum_{i=1}^{\ed}\theta_in_i$. In analogy with the notation used for $\pmb{\theta}$, we can also write $\pmb{n}=\{\vec{n}_1,\dots,\vec{n}_r\}$ with $\vec{n}_\ell\equiv \{n_{\ell,1},\dots,n_{\ell,d_G}\}$ and $\{n_{\ell,1},\dots,n_{\ell,d_G}\}\to \{n_{\ell d_G+1},\dots, n_{(\ell+1)d_G}\}$, emphasizing which components are associated to different copies of the group $G$. Since for the long-wavelength behavior of the system one can restrict to small momenta, we can expand the plane waves on the product domain to second order in the momenta $(\vfm,\pmb{n})$. In this way we obtain, for the correlator\footnote{Here we are choosing a normalization for the Casimir operator such that the kinetic kernel in the non-compact case of large $a$ matches with the one for $\R$, as we will see below.} in~\eqref{Fouriercoefficients},
\begin{equation}
    \hat{C}(\vfm,\pmb{n})\approx \int\text{d}^\ed \theta\text{d}^\std\phi
    \left\{1-\frac{1}{2}\left[(\pmb{\phi}\cdot \vfm)^2+\left(\frac{\pmb{\theta}\cdot\pmb{n}}{\rvp}\right)^2\right]\right\}
    C(\pmb{\phi},\pmb{\theta}),
\end{equation}
wherein the first order term vanishes due to isotropy (both in 
local and non-local variables). Similar symmetry arguments allow us to conclude that, up to second order\footnote{Notice that while in general there may be geometric anisotropies introduced by the non-locality of the interactions, they are unimportant at this perturbative order, as an expansion of~\eqref{Fouriercoefficients} at second order in powers of the \qmark{momenta} shows explicitly.} in $(\vfm,\pmb{n})$,
\begin{equation}\label{eqn:expcorrelation}
    \frac{\hat{C}(\vfm,\pmb{n})}{\hat{C}(\pmb{0},\pmb{0})}\approx \left\{1-\frac{1}{\hat{C}(\pmb{0},\pmb{0})}\left[\frac{\fm^2}{2n} \int\text{d}^{\ed} \theta\text{d}^\std\phi\,\phi^2C(\pmb{\phi},\pmb{\theta})+\frac{n^2}{2\ed\rvp^2} \int\text{d}^{\ed}\theta\text{d}^\std\phi\,\theta^2C(\pmb{\phi},\pmb{\theta})\right]\right\},
\end{equation}
where $\fm^2\equiv \vfm\cdot\vfm$, $\phi^2\equiv \pmb{\phi}\cdot\pmb{\phi}$, and similarly for $n^2$ and $\theta^2$. From this expression we can immediately identify the (modulus of the) correlation lengths in the local frame-variable and non-local geometric directions
\begin{subequations}\label{eqn:separatecorr}
\begin{align}\label{eqn:mattercontr}
    \xiloc^2&\equiv \frac{1}{2\std\hat{C}(\pmb{0},\pmb{0})}\int\text{d}^{\ed} \theta\text{d}^\std\phi\,\phi^2C(\pmb{\phi},\pmb{\theta})\,,\\\label{eqn:geocontr}
    \xinloc^2&\equiv \frac{1}{2\ed\hat{C}(\pmb{0},\pmb{0})}\int\text{d}^{\ed} \theta\text{d}^\std\phi\,\theta^2C(\pmb{\phi},\pmb{\theta}),
\end{align}
\end{subequations}
wherein the zero-mode of the two-point function in Fourier space is given by
\begin{equation}
    \hat{C}(\pmb{0},\pmb{0})=\int_{D}\diff^\std \phi\diff^\ed\theta\,C(\pmb{\phi},\pmb{\theta})= \frac{1}{b_{\pmb{0}}}\,.
\end{equation}

When the group $G$ is non-Abelian, one could in principle follow the same derivation used here, for instance by replacing the Fourier transform employed in~\eqref{eqn:firststepderivation} with a non-commutative one~\cite{Guedes:2013vi}. In practice, this would amount to an expansion in terms of non-commutative plane waves
\begin{equation}
    e_\star^{ik(g)\cdot X}\equiv \sum_{n=0}^\infty\frac{i^n}{n!}k(g)^{i_1}\dots k(g)^{i_n}X_{i_1}\star\dots\star X_{i_n}\,,
\end{equation}
where $X\in\mathfrak{g}$ and\footnote{Notice that by assumption the Lie group is taken to be exponential.} $k(g)=-i\log g$, up to the second order in $X$. However, even at this order, possibly infinitely many powers of the group variable would have to be included~\cite{Guedes:2013vi}, meaning that one would not be able to write the geometric contribution to the correlation length as a second order moment of the correlation function, as it is the case for~\eqref{eqn:geocontr}. However, under some approximations (e.g., semi-classicality, see~\cite{Li:2017uao} for an example) one can effectively treat the non-commutative star product as a standard point-wise one, thus solving the aforementioned issue. This point will be discussed in more detail in a forthcoming paper~\cite{Marchetti:2021xyz}.

Let us now use~\eqref{eqn:mattercontr} and~\eqref{eqn:geocontr} to explicitly compute the local frame-variable and non-local geometric contributions with finite $a$. 
\paragraph{Local (frame-)variable contribution.}
By writing the two-point function in Fourier space and integrating over the group variables we find
\begin{equation}
    \xiloc^2\equiv \frac{b_{\pmb{0}}}{2n}\sum_{i=1}^{\std}\int\frac{\diff^\std\fm}{(2\pi)^\std}\frac{\diff^\std\phi~ \phi_i^2}{\cc(\pmb{0})\sum_b 
    \fm_b^2+b_{\pmb{0}}}e^{i\pmb{\phi}\cdot\vfm}\,.
\end{equation}
By performing the integration over all the $\phi_b$ with $b\neq i$, we obtain $\std-1$ delta functions $\delta(p_b)$, which are absorbed by $\std-1$ integrals over $\diff p$. As a result, defining $m^2(\pmb{0})=b_{\pmb{0}}/\cc(\pmb{0})
$ we find
\begin{align}\label{eqn:mattercontribution}
     \xiloc^2&\equiv \frac{b_{\pmb{0}}}{2n}\frac{1}{\cc(\pmb{0})}\sum_{i=1}^{\std} \int\diff\phi_i \frac{\diff \fm_i}{2\pi}\frac{\phi_i^2}{\fm_i^2+m^2 (\pmb{0})}e^{i\fm_a\phi_a}\nonumber\\
     &=\frac{b_{\pmb{0}}}{4n}\frac{1}{\cc(\pmb{0})\vert m(\pmb{0})\vert}\sum_{i=1}^{\std}\int\diff\phi_i\phi_i^2 e^{-\vert\phi_i\vert \vert m (\pmb{0})\vert}=\frac{b_{\pmb{0}}}{\cc(\pmb{0})m^4(\pmb{0})}=\frac{\cc(\pmb{0})}{b_{\pmb{0}}}\,.
\end{align}
Notice that the strength of the minimal coupling of the local degrees of freedom to the non-local geometric ones is governed by the $\alpha$-factor which can be absorbed in $\xiloc$.

\paragraph{Non-local (geometric) contribution.}

Let us now move to the computation of the geometric contribution to the correlation length in the compact case (finite $\rv$). Starting from the definition in~\eqref{eqn:geocontr}, we can write 
\begin{equation}
    \xinloc^2=\frac{b_{\pmb{0}}}{2\ed}\int\diff^\ed\theta\left(\sum_{\ell=1}^\ed\theta_{\ell}^2\right) \sum_{\pmb{n}}\frac{1}{\rv^\ed}\frac{1}{\sum_\ell n_\ell^2/\rvp^2+b_{\pmb{n}}}e^{i\pmb{n}\cdot\pmb{\theta}/\rvp} \quad .
\end{equation}
By performing the integration over all $\theta_{\ell'}$ with $\ell'\neq \ell$, we obtain
\begin{equation}
    \xinloc^2=\frac{b_{\pmb{0}}}{2\ed}\sum_\ell\sum_{n_\ell}\frac{1}{\rv}\frac{1}{\sum_\ell n_\ell^2/\rvp^2+b_{n_\ell}}\int_{-\pi \rvp}^{\pi \rvp}\diff\theta_
    \ell\,\theta_{\ell}^2 e^{i\theta_\ell n_\ell/\rvp}\,,
\end{equation}
where $b_{n_\ell}$ is given by $b_{\pmb{n}}$ where all $n_{\ell'}=0$ except for $\ell=\ell'$, which is still unconstrained. Let us consider $n_\ell=0$ and $n_\ell\neq 0$ separately, i.e.
\begin{align}\label{eqn:xiggeneralcompact}
    \xinloc^2&=\frac{b_{\pmb{0}}}{4\pi\ed}\sum_\ell\left\{\frac{2\pi^3\rvp^2}{3b_{\pmb{0}}}+\sum_{n_\ell\neq 0}\frac{1}{n_\ell^2/\rvp^2+b_{n_\ell}}\int\frac{\diff\theta_\ell}{a}\theta_\ell^2e^{in_\ell\theta_\ell/\rvp}\right\}\\
    &=\frac{b_{\pmb{0}}}{4\pi\ed}\sum_\ell\left\{\frac{2\pi^3\rvp^2}{3b_{\pmb{0}}}+\sum_{n_\ell\neq 0}\frac{\rvp^2}{n_\ell^2}\frac{4\pi(-1)^{n_\ell}}{n_\ell^2/\rvp^2+b_{n_\ell}}\right\}.\nonumber
\end{align}
Now, \emph{for any finite $a$} we can take the limit of small $\mu$ (characterizing the phase transition) in the denominator of the second term in curly brackets to obtain, at first order in $\mu$,
\begin{equation}\label{eqn:correlationlengthcompact}
    \xinloc^2\simeq\frac{\pi^2\rvp^2}{2}\left[\frac{1}{3}- \frac{7\pi^2b_{\pmb{0}}\rvp^2}{180}\right].
\end{equation}
So, for $\mu\to 0$, the geometric contribution to the correlation length is only given by $\xinloc^2=\pi^2\rvp^2/6=a^2/24$, which is finite and therefore negligible with respect to the local frame-variable contribution, which is instead diverging as $\mu\to 0$. 
We may take the fact that $\xinloc<a<\infty$ as an indication that for the compact domain (neglecting the local coordinates) there is no phase transition, as expected.

\subsection{Non-compact case}\label{subsec:explicitexamplescorr}
In order to address the non-compact case, we will consider the large $a$ regime (or, equivalently, large $\tilde{a}$) where appropriate. Notice that this means \qmark{decompactifying uniformly} all the $\text{U}(1)$ factors in $G$. Of course, a general non-compact connected Abelian Lie group $G$ may not be in this form. However, it can be shown that it must be isomorphic to $\text{U}(1)^t\times F$, with $F$ a vector space and $t$ a positive integer~\cite{ref1}. So, as we will see below, it is easy to draw general conclusions from the results below and those obtained in the above section.

Notice that in the non-compact case $G\cong \mathbb{R}^{d_G}$, local frame and non-local geometric variables are treated kinematically on the same footing, so it would make sense to define, following the same procedure that led us to~\eqref{eqn:expcorrelation}, a \qmark{total} correlation length given by
\begin{equation}\label{eq:correlationlengthtotal}
    \xi^2\equiv \frac{1}{2(\std+\ed)\hat{C}(\pmb{0},\pmb{0})}\int_D\diff^\ed \theta\diff^\std\phi \Vert (\pmb{\theta},\pmb{\phi})\Vert^2_D\,C(\pmb{\phi},\pmb{\theta})\,,
\end{equation}
where $\Vert (\pmb{\theta},\pmb{\phi})\Vert^2_D$ is the distance on $D\cong \mathbb{R}^{\std+\ed}$ defined by the product topology. However, since local and non-local degrees of freedom are treated differently at the dynamical level (entering respectively in a local and non-local manner in the interactions), in the following it will be more appropriate to deal with $\xiloc^2$ and $\xinloc^2$ separately, as defined by~\eqref{eqn:separatecorr}. In particular, notice that by~\eqref{eqn:mattercontr} the local frame-variable contribution will be still given by~\eqref{eqn:mattercontribution} even when we consider the non-compact group case.

Strictly speaking, therefore, we will only need to compute $\xinloc^2$. Before doing this, however, it is instructive to rewrite explicitly the correlation function in group space. Since the right-hand-side of~\eqref{eqn:geocontr} involves an integral over the local frame variables, it will be enough to consider the correlation function of a model without local directions.

\paragraph{Non-local (geometry) 
correlation function in group space.}
From~\eqref{Fouriercoefficients}, we deduce that the correlation function for interactions of order $\ooi$ is, in this context, and 
using the notation introduced at the beginning of Section \ref{subsec:compactcase},

\begin{equation}\label{eqn:corrfunctiongroup}
     C(\pmb{\theta})=\frac{1}{\cs^r}\sum_{\pmb{n}}\hat{C}(\pmb{n})e^{i\pmb{n}\cdot\pmb{\theta}/\rvp}\,,
\end{equation}
where
\begin{equation}
    \hat{C}(\pmb{n})=\frac{1}{\frac1{\rvp^2}\sum_{\ell=1}^{\ed} n_\ell^2 + \mu - {\mu}\sum_\gamma{\ct_\gamma}\nloph_\gamma(\pmb{n})}
    \equiv \frac{1}{\frac{1}{\rvp^2}\sum_{\ell=1}^{\ed} n_\ell^2 + b_{\pmb{n}}} \, ,
\end{equation}
and where $b_{\pmb{n}}$ is given here by 
\begin{equation}\label{eqn:bncoefficientsexplicit}
    b_{\pmb{n}}=\mu\left[1-\sum_{\gamma;\nv_\gamma=\ooi} \ct_\gamma \sum_{p=0}^\rk \sum_{(\ell_0,...,\ell_p)} \nlc_{\ell_0...\ell_p} \prod_{\ell=\ell_1}^{\ell_p} \delta_{\vec{n}_{\ell,0}}\right],\quad\delta_{\vec{n}_\ell,0}\equiv \prod_{i=1}^{d_G}\delta_{n_{\ell,i},0}\,.
\end{equation}
Given this form of the $b_{\pmb{n}}$ coefficients, it is useful to decompose the sums in~\eqref{eqn:corrfunctiongroup} in different combinations of zero-modes

\begin{align}\label{eqn:decompositioncorrelation}
    C(\pmb{\theta})&= \frac{1}{\cs^r} \sum_{s=0}^\rk \sum_{(\ell_1,...,\ell_s)} \sum_{\substack{\vec{n}_{\ell_{s+1}},...,\vec{n}_{\ell_\rk}\neq 0\\\vec{n}_{\ell_{1}},...,\vec{n}_{\ell_{s}}=0}}\hat{C}(\pmb{n})e^{i\pmb{n}\cdot\pmb{\theta}/\rvp}
    \\
    &= \frac{1}{\cs^r} \left( \sum_{\pmb{n}\neq \pmb{0}}\hat{C}(\pmb{n})e^{i\pmb{n}\cdot\pmb{\theta}/\rvp} 
    + \sum_{\ell=1}^r\delta_{\vec{n}_\ell,0} \sum_{\{\pmb{n}\}\backslash\{\vec{n}_\ell\}\neq \pmb{0}}\hat{C}(\pmb{n})e^{i\pmb{n}\cdot\pmb{\theta}/\rvp}\nonumber
    + \dots + \hat{C}(\pmb{0}) \right)\,.\nonumber
\end{align}
We call a contribution in which $s$ arguments $\vec{n}_{\ell_{1}},...,\vec{n}_{\ell_{s}}$ are zero a \emph{$s$-fold zero-mode} and denote $b_{\ell_1,...,\ell_s}$ the evaluation of the effective mass $b_{\pmb{n}}$ on this mode.
Note that this effective mass $b_{\ell_1,...,\ell_s}$ is not only different for each single term in the sum over products of Kronecker deltas   $\sum_\gamma{\ct_\gamma}\nloph_\gamma(\pmb{n})$ but also for any combination of them.
For example, if $\sum_\gamma$ sums over quartic melonic interactions of each colour $c=1,...,\rk$ (Tab. \ref{tab:vertexexamples}), 
already the single-delta terms $\delta_{\vec{n}_c,0}$ give rise to any combination of single zeros to multiple zero-modes in the correlation function.
Furthermore, an $s$-fold zero-mode contribution only depends on the remaining $\rk-s$ variables not only in momentum but also in position space,

\begin{equation}
    C_s(\vec{\theta}_{c_1},\dots,\vec{\theta}_{c_{r-s}})=\frac{1}{\rv^{\gd s}}\frac{1}{\rv^{\gd(\rk-s)}} \sum_{\substack{\vec{n}_{c_m}\,\neq\, \vec{0}\\[0.5mm]
    \forall m\,=\,1,\,\dots\,,\,\rk-s}}
    \hat{C}_s(\vec{n}_{c_1},\dots,\vec{n}_{c_{r-s}})e^{i(\vec{n}_{c_1}\cdot\vec{\theta}_{c_1}+\dots +\vec{n}_{c_{r-s}}\cdot\vec{\theta}_{c_{r-s}})/\rvp},
\end{equation}
where $\vec{n}_\ell\cdot\vec{\theta}_\ell=\sum_{i=1}^{d_G}n_{\ell,i}\theta_{\ell,i}$ and  $\hat{C}_s(\vec{n}_{c_1},\dots,\vec{n}_{c_{r-s}})\equiv \hat{C}(\pmb{n})\vert_{\vec{n}_{d_1}=\dots=\vec{n}_{d_{s}}=\vec{0}}$.

Let us now consider the non-compact limit of large~$a$. In this limit, we denote $n_\ell/\rvp\equiv p_\ell$, with $p_\ell\in \mathbb{R}$, and the discrete sums become 
$\sum_{n_\ell}/\rv\to \int\diff p_\ell/(2\pi)$. In this limit, the contribution of an $s$-fold zero-mode to the correlation function is

\begin{equation}
    C_s(\vec{\theta}_{c_1},\dots,\vec{\theta}_{c_{r-s}})
    = \frac{1}{\rv^{\gd s}}\int\frac{\diff^{\gd(\rk-s)}p}{(2\pi)^{\gd(\rk-s)}}\frac{e^{i\pmb{p}_{\rk-s}\cdot\pmb{\theta}_{\rk-s}}}{p^2_{\rk-s}+b_{\ell_1,...,\ell_s}}\,,
\end{equation}
where the subscript $\rk-s$ in the above quantities are to remind that they can be seen as vectors in a $\gd(\rk-s)$ dimensional space. The result of the  integration depends of course critically on the sign of $b_{\ell_1,...,\ell_s}$. If it is positive, this contribution shows an asymptotic exponential decay with a cut-off scale given by $1/\sqrt{b_{\ell_1,...,\ell_s}}$. 
On the other hand, when $b_{\ell_1,...,\ell_s}$ is negative, after an appropriate regularization, one obtains a polynomially suppressed oscillating contribution, as one can see from the explicit computations in Appendix~\ref{app:integrals}. While exponentially decaying correlations are typically encountered in local statistical field theories, this oscillating behavior produces long-range correlations (see also the discussion below) which are however not associated with the phase transition characterized by $\mu\to 0$. Such contributions are indeed produced when the interactions included in the action generate only $\gd s'$ with $s'>s$ zero-modes in \eqref{eqn:decompositioncorrelation}.

\paragraph{Non-local (geometry) contribution to the correlation length.}

Since in this case we are interested in taking the large $a$ limit 
\emph{before} taking the $\mu\to 0$ limit, the procedure employed in Section \ref{subsec:compactcase} is not adequate anymore (recall that in~\eqref{eqn:correlationlengthcompact} we have neglected higher powers in $a$). Moreover, it is useful to explicitly employ the decomposition in \eqref{eqn:decompositioncorrelation} in order to better understand how different $s$-fold zero-modes contribute to the correlation length. By definition \eqref{eqn:geocontr}, we can write the correlation length as

\begin{align}
    \xinloc^2&=\frac{b_{\pmb{0}}}{2\ed}\int\diff^{\ed}\theta\,\theta^2\frac{1}{\cs^\rk} \sum_{s=0}^\rk \sum_{(\ell_1,...,\ell_s)} 
    \sum_{\substack{\vec{n}_{\ell_{s+1}},...,\vec{n}_{\ell_\rk}\neq 0\\\vec{n}_{\ell_{1}},...,\vec{n}_{\ell_{s}}=0}}\hat{C}(\pmb{n})e^{i\pmb{n}\cdot\pmb{\theta}/\rvp}\nonumber\\
    &=\frac{b_{\pmb{0}}}{2\ed}\sum_{s=0}^\rk \sum_{(\ell_0,...,\ell_s)} 
   \frac{1}{\cs^{\rk-s}} \sum_{\substack{\vec{n}_{\ell_{s+1}},...,\vec{n}_{\ell_\rk}\neq 0\\\vec{n}_{\ell_{1}},...,\vec{n}_{\ell_{s}}=0}}\hat{C}(\pmb{n})\int \frac{\diff^{\gd s}\theta}{\cs^{s}}\int\diff^{\gd(\rk-s)}\theta\,\theta^2 e^{i\pmb{n}\cdot\pmb{\theta}/\rvp}
\end{align}
Let us now split up the sum in $\theta^2$ into two contributions $\theta^2=\theta_s^2+\theta_{\rk-s}^2$, being associated to the modes $\vec{n}_{\ell_{1}},...,\vec{n}_{\ell_{s}}$ and  $\vec{n}_{\ell_{s+1}},...,\vec{n}_{\ell_\rk}$, respectively. By construction, in the above equation $e^{i\pmb{n}\cdot\pmb{\theta}/\rvp}=e^{i\pmb{n}_{\rk-s}\cdot\pmb{\theta}_{\rk-s}/\rvp}$ only depends on these latter $\rk-s$ variables. Hence, the integral
 
\begin{equation}
    \int \frac{\diff^{\gd s}\theta}{\cs^{s}}\int\diff^{\gd(\rk-s)}\theta~\theta^2_s e^{i\pmb{n}\cdot\pmb{\theta}/\rvp}=\int \frac{\diff^{\gd s}\theta}{\cs^{s}}\theta_s^2\int\diff^{\gd(\rk-s)} \theta\,e^{i\pmb{n}\cdot\pmb{\theta}/\rvp}
\end{equation}
vanishes for each $s<r$ because the integration over the $\rk-s$ variables produces delta functions over the corresponding momenta, which are, however, by construction different from zero. On the other hand, for $s=r$, we only have one contribution, proportional to $a^2$. This is somehow expected since the non-compactness scale $a$ was introduced in order to tame the divergences associated to non-locality and the mean-field uniform solution. The quantity  $\xi^2$ can be seen in principle as proportional to both the scales of the theory: the physical one $\mu^{-1}$ and the regulator $a^2$, having the same \lq dimension\rq. We argue that the divergence $\sim a^2$ of the correlation length is unphysical and that, similarly to what is done in the usual renormalization procedure of dimensionful quantities in local field theories, it should be subtracted in order to obtain the physical correlation length~\cite{Salmhofer:1999uq,zinn2007phase}.

Let us therefore consider the remaining contributions. The integration over $\diff^{\gd s}\theta$ cancels with $\cs^s$, so the only non-trivial integral involves the remaining $\rk-s$ variables. This is easily done in the limit of very large $\cs$. In this case, the sum over the non-zero momenta divided by $\cs^{\rk-s}$ turns into an integral, and we have to compute
\begin{equation}\label{eqn:auxintegral1}
    \int\frac{\diff^{\gd(\rk-s)}p}{(2\pi)^{\gd(\rk-s)}}\frac{1}{p^2_{\rk-s}+b_{\ell_1,...,\ell_s}}\int\diff^{\gd(\rk-s)} \theta\,\theta^2_{\rk-s}e^{i\pmb{p}_{\rk-s}\cdot\pmb{\theta}_{\rk-s}}\,,
\end{equation}
where we have denoted, as before, $n_\ell/\rvp\equiv p_\ell$, with $p_\ell\in \mathbb{R}$, and\footnote{As in the previous paragraph, the point $p_\ell=0$ has also be added to the domain of integration. We remark again that, being the integrand regular at $p_\ell=0$ (for finite $\mu<0$), and $p_\ell=0$ being zero measure, this does not change the result.} $\sum_{n_\ell}/a\to \int\diff p_\ell/(2\pi)$.
For a negative effective mass $b_{\ell_1,...,\ell_s}<0$, the integral diverges. In this sense, all the $s$-fold zero-modes with a negative effective mass produce an infinite correlation length. This is expected since we have seen above that they generate oscillating correlations suppressed only by a power-law. This behavior is indeed indicative of correlations at any scale, regardless of the precise value of $\mu$, taken here to be finite. Here, we are however interested in a finite correlation length, diverging only when the critical point is reached (i.e.\ when $\mu\to 0$), in terms of which one can interpret the behavior of the system around the phase transition. For this purpose, it is therefore natural to just not consider the terms with negative effective mass. From the structure of the interaction terms and the arguments discussed below equation \eqref{eqn:decompositioncorrelation}, it is clear that a negative effective mass can be obtained only from $s$-fold zero-modes with $s<s_0$, where $\gd s_0$ is the minimum number of delta functions appearing in the interactions. For example, a multi-trace containing a fundamental melon (two vertices connected by $\rk$ edges) has $s_0=0$, a melonic interaction $s_0=1$, a necklace interaction $s_0=\rk/2$ and a simpicial interaction $s=\rk-1$, see Table \ref{tab:vertexexamples} for examples. In practice, this means that we need to compute the contribution to the correlation length coming from $s\ge s_0$-fold zero-modes.

\eqref{eqn:auxintegral1} can be easily computed, for a positive effective mass. It is just
\begin{equation}
    \sum_{l}\int\frac{\diff p}{(2\pi)}\frac{1}{p^2_{\ell_l}+b_{\ell_1,...,\ell_s}}\int\diff\theta\,\theta^2_{\ell_l}e^{i p_{\ell_l}\theta_{\ell_l}}=\frac{2\gd (\rk-s_0)}{b_{\ell_1,...,\ell_s}^2}\,,
\end{equation}
as one can show for instance by exchanging the integration order and writing $\theta^2_{\ell_a}$ in terms of derivatives acting on the exponential function. As a result, we finally obtain
\begin{equation}\label{eqn:xigcont}
    \xinloc^2=\sum_{s=s_0}^r\frac{\gd(\rk-s_0)}{\ed}\sum_{(\ell_1,...,\ell_s)} \frac{b_{\pmb{0}}}{b_{\ell_1,...,\ell_s}^2}\,.
\end{equation}

In the limit $\mu\to 0$, the correlation length $\xinloc^2$ diverges as $\mu^{-1}$, exactly as in the local, non-compact case, so this result for the correlation length qualitatively agrees with the standard result for a local statistical field theory on $\mathbb{R}^r$, though being quantitatively different.
The above result also clarifies what happens in the case of a non-compact group of the form $\text{U}(1)^t\times F$. The vector space part would in fact contribute to the correlation length as computed here, while the compact part would contribute as computed in equation~\eqref{eqn:correlationlengthcompact}. As a result, therefore, the contribution to the correlation length coming from the compact directions would be negligible, even in absence of local frame variables. 

\section{Ginzburg criterion}\label{sec:ginzburg}

The Ginzburg criterion is a way to test the reliability of mean-field theory by checking whether fluctuations remain small. Concretely, for mean-field theory to be self-consistent, it requires that fluctuations of the order parameter $\Phi$ \emph{averaged on an appropriate region} $\Omega$ should be much smaller then the value of the mean order parameter $\Phi_0$ itself averaged on such region, i.e.,~\cite{nielsen}

\begin{equation}\label{eq:Ginzburgmotivation}
    \left\langle(\delta\Phi)^2\right\rangle_\Omega\ll \left\langle\Phi_0^2\right\rangle_{\Omega}\,.
\end{equation}
For the Ginzburg criterion, the averaging region is crucial. For applications to three-dimensional statistical systems, one typically chooses $\Omega\equiv \Omega_\xi\sim \xi^3$ where $\xi$ is the correlation length.
The reason is that in a system where correlations are relevant only until distances of order $\xi$, regions of linear size $\Omega_\xi$ are practically statistically independent~\cite{Kleinert:2001ax}. However, this is not the case in general. In particular, consider an anisotropic system, having two different correlation lengths, say $\xi_\perp\equiv \xi$ and $\xi_\parallel\equiv f(\xi)$. In this case, one cannot just choose $\Omega_\xi\sim \xi^3$, because the system shows different correlation properties in different directions. An explicit example of this issue is provided by an Ising ferromagnect where the magnetic (electric) dipole moments are only coupled by dipolar interactions. In such a case, one has two correlation lenghts: $\xi_\parallel\sim \xi^2$, and $\xi_\perp\sim \xi$. The averaging region should then be chosen as $\Omega_\xi\sim \xi^4$, eventually leading to an \qmark{almost mean-field behavior}~\cite{nielsen}.

From these general arguments we can see that some care should be taken when we try to concretely evaluate the Ginzburg criterion, since our model clearly displays anisotropy between the local and non-local sector.
As we are able to identify the correlation lengths in group space and in reference frames space via the second-moment method described in the previous Section, we can define the region $\Omega_\xi$ as 
\begin{equation}\label{eqn:integrationregion}
    \Omega_\xi\sim \xiloc^\std\times \xinloc^\ed\,,
\end{equation}
where, as before, we are considering $\std$ local frame variables and $\ed=rd_G$ non-local variables on $\rk$ copies of  $G\cong \text{U}(1)^{d_G}$ and we have assumed isotropy on the local frame-variable and non-local geometric directions separately.

Since fluctuations in the order parameter are captured by the correlation function, we see that by defining a quotient measuring their relative importance, the so-called $Q$ parameter, as
\begin{equation}
    Q\equiv  \frac{\int_{\Omega_\xi} {\diff}^{r}g~{\diff}^\std\ff~C(\pmb{g}
    , \pmb{\ff}
    )}{\int_{\Omega_\xi}{\diff}^{r}g{\diff}^\std\ff~ \gf_0^2}
    = \frac{\int_{\Omega_\xi} {\diff}^{\ed}\theta~{\diff}^\std\ff~C(\pmb{\theta}
    , \pmb{\ff}
    )}{\int_{\Omega_\xi}{\diff}^{\ed}\theta~{\diff}^\std\ff~ \gf_0^2}\,,
\end{equation}
condition~\eqref{eq:Ginzburgmotivation} is equivalent to $\vert Q\vert\ll 1$. Note that the second equation holds in the specific Abelian case we are considering (where $\Omega_\xi$ is given by~\eqref{eqn:integrationregion}). Here we will be interested in studying the value of $Q$ at the phase transition characterized by $\mu\to 0$,
in order to assess the validity of mean-field methods around the phase transition. If $Q\gg 1$ the system is strongly interacting and fluctuations are large, mean-field theory and the Gaussian approximation are insufficient to give a description of the phase transition.

We will employ the working assumptions of Section \ref{sec:correlationlength}, and in particular we will consider $G\cong \text{U}(1)^{d_G}$, with each $\text{U}(1)$ volume given by $a$. 
First, in Section \ref{subsec:ginzburgnoncompact}, we will consider only the non-local variables in the large $a$ (non-compact) limit.\footnote{In Section \ref{subsec:compactcase}, we have seen that the correlation length is always finite in the compact case, even when $\mu\to 0$, suggesting that the phase transition is not present in these models, as also emphasized in~\cite{Pithis:2018bw}.} Then, in Section \ref{subsec:ginzburgmatter}, we will compute it in the case in which local degrees of freedom are included.

\subsection{Ginzburg criterion for non-local variables in the non-compact limit}\label{subsec:ginzburgnoncompact}
 
As we have seen in Section~\ref{subsec:explicitexamplescorr}, the correlation function of this theory is characterized, due to its non-locality, by different contributions, characterized by their number $s$ of zero-modes. These terms behave quite differently depending on whether the effective mass term $b_{\pmb{n}}$,~\eqref{eq:effectivemass}, is positive or not. Contributions with negative effective mass have to be excluded since they do not play a role in the phase transition. This is because the relative $C_s$ show no cut-off of the correlations at large scales, even far from criticality. We thus suggest to exclude these long range contributions from the computation of the $Q$~integral, since, as we have reviewed above, the Ginzburg criterion is directly related to the exponential fall off of correlations beyond a certain scale, i.e. the correlation length.  

Having made these premises, let us explicitly compute the $Q$ integral. The denominator of the integral for a sum over quartic interactions with uniform minimum~\eqref{eq:minimum} is

\begin{equation}\label{eqn:denominatornonlocal}
    \int_{\Omega_\xi}{\diff}^{r}g~\gf_0^2
    = \xi^\ed \gf_0^2
    = -\frac{\mu}{4\sum_\gamma\lambda_{\gamma}}\left(\frac{\xi}{a}\right)^\ed\,,
\end{equation}
where $\xi\equiv \xinloc$ is given by~\eqref{eqn:xigcont}. 
Thus $\xi^2\sim \mu^{-1}$ asymptotically. 
On the other hand, the numerator is a sum of integrals over terms $\hat{C}_s$ with $s$ zero-modes,~\eqref{eqn:decompositioncorrelation}, contributing

\begin{equation}\label{eqn:sumovercontributions}
    \sum_{s = s_0}^r\left(\frac{\xi}{\rv}\right)^{d_G s}\frac{1}{\rv^{\gd(\rk-s)}}\int\diff^{\gd(\rk-s)}\theta
    \sum_{(\ell_1,...,\ell_s)} \sum_{\vec{n}_{c_{s+1}},\dots,\vec{n}_{c_{\rk}}} \frac{1}{\frac1{\rvp^2}\sum\limits_{p=s+1}^\rk \vec{n}_{\ell_p}^2 + b_{c_1...c_s}} e^{i\sum\limits_{p=s+1}^\rk \vec{n}_{\ell_p} \cdot \vec{\theta}_{\ell_p} / \rvp}
\end{equation}
where $s_0$ is the minimal number of zero-modes depending on the specific non-local interaction, see Section~\ref{subsec:explicitexamplescorr}.

Integrating first over $\theta$, we obtain $d_G(r-s)$ Kronecker deltas multiplied by $a$, which fix each $\vec{n}_{\ell_p}$ to zero and cancel the prefactor $a^{d_G(r-s)}$. As a result, we find

\begin{equation}\label{eqn:qnnoncompact}
    \int_{\Omega_\xi} {\diff}^{\ed}\theta~C(\pmb{\theta})
    = \sum_{s = s_0}^\rk \left(\frac{\xi}{\rv}\right)^{d_G s}
    \sum_{(\ell_1,...,\ell_s)} \frac{1}{b_{c_1...c_s}}
    \,.
\end{equation}

Taking everything together, we find for a sum over quartic interactions $\gamma$ the result 
\[
Q = -\frac{4\sum_\gamma\lambda_\gamma}{\mu^2} \sum_{s=s_{0}}^\rk f_s \left(\frac{\xi}{\rv}\right)^{d_G (s-\rk)}
\]
where the sum $f_s := \mu\sum{1}/{b_{c_1...c_s}}$ over all $s$-fold zero-modes $(\ell_1,...,\ell_s)$ is independent of $\mu$  since any $b_{c_1...c_s}$ is proportional to $\mu$.

The asymptotics of this sum now depends crucially on whether the size of compactness~$a$ is finite or taken to be arbitrarily large.
If $a$ is kept finite one has large-$\xi$ asymptotics $Q\sim-4f_\rk\sum_\gamma\lambda_\gamma/\mu^2$ which diverges due to $\mu\sim\xi^{-2}$ and corresponds to the usual result for a zero-dimensional field theory as previously found \cite{Pithis:2018bw} and known in general for fields on a compact domain \cite{Benedetti1403}.
On the other hand, if one takes the large $a$ limit one recovers exactly the least dominant term in $\xi$ since $a$ occurs only in the combination $\xi/a$, i.e.

\begin{equation}\label{eq:Qgeometric}
    Q \underset{a\to\infty}{\sim}\frac{-4f_{s_0}\lambda_{\gamma}}{\mu^2}\left(\frac{\xi}{\rv}\right)^{\gd(s_0-\rk)}
    \underset{\xi\to\infty}{\sim} -4f_{s_0}\lambda_{\gamma}\frac{\xi^{4-\gd(\rk-s_0)}}{\rv^{\gd(s_0-\rk)}}.
\end{equation} 
Therefore, in this limit there is a critical rank given by $\rk_c=s_0+4/\gd$: While for $\rk<\rk_c$ the quotient $Q$ diverges and the theory has no phase transition in the Gaussian approximation, for $\rk>\rk_c$ the $Q$ is small, the Gaussian approaximation holds and there is thus a phase transition which can be described by mean-field theory.
Notice that $s_0=0$ for disconnected (multi-trace) interactions containing a fundamental melon and therefore $Q$ has $\xi^{4-\ed}$ asymptotics which characterizes local field theories in $d=\ed$ dimensions.
This is because such non-local multi-trace interactions can be understood as an interaction of a local vector theory, see e.g.~\cite{Rivasseau:2015im}. 
On the other hand, when they are absent and melonic interactions are present, $Q\sim\xi^{4-d_G(\rk-1)},$ as expected from FRG analysis, see for instance~\cite{Pithis:2020sxm,Pithis:2020kio}. 
In particular, this means that in this case the critical rank $\rk_c$ for $d_G=1$ is $5$. In general, interactions in the Landau-Ginzburg setting are chosen \enquote{by hand} and there is no reason not to leave out any interactions.
One can equally well restrict the action to any subdominant interactions as for example necklace interactions with $s_0=\rk/2$.
The question whether such action is stable along all scales can only be answered by methods which fully take into account the renormalization group flow.

Quartic interactions are the most relevant ones for testing the validity of mean-field phase transitions but our setting allows to compute $Q$ also for interactions of any other order.
For a single interaction given by a $\nv_\gamma$-valent graph $\gamma$ yielding at least an $s_0$-fold zero-mode and by using ~\eqref{eq:nonzerosolution} for the general vacuum solution $\gf_0$, one finds in the large $a$ limit
\begin{align}\label{eq:GinzburgQgeneralscaling}
    Q \sim \frac{\left(\frac{\xi}{a}\right)^{s_0\gd}{\mu}^{-1}}{\left(\frac{\xi}{a}\right)^{\rk\,\gd}\lambda_\gamma^{\frac{\nv_\gamma-2}{2}} \mu^{\frac{2}{\nv_\gamma-2}}}
&= \lambda_\gamma^{\frac{2}{\nv_\gamma-2}} \mu^{-\frac{\nv_\gamma}{\nv_\gamma-2}} \left(\frac{\xi}{a}\right)^{-(\rk-s_0)\gd}\nonumber \\
&\underset{\mu\sim\frac{1}{\xi^{2}}}{=} 
\left(\lambda_\gamma \, \xi^{\nv_\gamma
- \left(\frac{\nv_\gamma}2-1\right)(\rk-s_0)\gd} \, a^{\left(\frac{\nv_\gamma}2-1\right)(\rk-s_0)\gd} \right)^{\frac{2}{\nv_\gamma-2}}
\end{align}
In fact, this is simply the well known scaling of the coupling in a momentum scale $k\sim1/\xi$ known from renormalization (see for example Eq.\ (3.1), (3.5) in~\cite{Pithis:2020kio}, there $n=\nv_\gamma/2$).
That is, the Ginzburg $Q$ at large $\xi$ in the large $a$ limit for a single interaction $\gamma$ is exactly (up to some irrelevant constant) some power of the rescaled coupling $\bar{\lambda}_\gamma$,
\[
Q \sim \lambda_{\gamma}^{\frac{2}{\nv_\gamma-2}} \frac{\xi^{\frac{2 V_{\gamma}}{V_{\gamma-2}}-d_G(r-s_0)}}{a^{d_G(s_0-r)}}= \bar{\lambda}_\gamma^{\frac{2}{\nv_\gamma-2}} \, .
\]
Thus, from the renormalization perspective this calculation of the Ginzburg $Q$ is just an alternative way to determine the scaling of couplings with an RG scale, or equivalently, the scaling exponents of the couplings at the Gaussian fixed point.

\paragraph{Imposing the closure constraint.}

As commented on in Section \ref{subsec:gaussianapprox}, the imposition of the closure condition leads to a slight modification of the correlator, giving~\eqref{Fouriercoefficientsgc}. To understand how the inclusion of the gauge constraint affects the numerator of the Ginzburg-parameter, one first has to appreciate that it essentially contaminates the expansion of the correlation function in~\eqref{eqn:corrfunctiongroup} with a factor $\delta(\sum_i n_i)$ via its Fourier coefficients and thus has only a minor impact on the decompositions of the different contributions in~\eqref{eqn:sumovercontributions}. When computing $Q$ for the middle terms in that decomposition, one notices that~\eqref{eqn:sumovercontributions} is modified by adding one more zero-mode, that is, the contamination through the Kronecker delta eliminates one sum over an $n$ which in turn also leads to another empty integral over the group yielding an additional factor in $\xi/a$. In~\eqref{eqn:sumovercontributions} this can simply be accounted for by setting either for the rank $r\to r-1$ or for the number of zero-modes $s\to s+1$. Likewise, these shifts are easily applied to the overall general Ginzburg parameter~\eqref{eq:GinzburgQgeneralscaling} to incorporate the influence of the gauge constraint. This is consistent with results on scaling dimensions obtained in renormalization studies of GFTs, see for instance~\cite{Benedetti:2016db,BenGeloun:2016kw} and in particular the discussion in Appendix C of~\cite{Pithis:2020kio}.

\subsection{Ginzburg criterion with local variables and non-local degrees of freedom}\label{subsec:ginzburgmatter}
Let us now consider the case in which both the local (frame) variables and the non-local (geometric) degrees of freedom are present. 

Using~\eqref{eqn:integrationregion}, the denominator is a simply generalization of~\eqref{eqn:denominatornonlocal},

\begin{equation}
    \int_{\Omega_\xi}{\diff}^{\ed}\theta~{\diff}^\std\ff~ \gf_0^2  \frac{-\mu}{\lambda_{\gamma}} \xiloc^\std\xinloc^\ed
    = -\frac{\mu}{4\sum_\gamma\lambda_{\gamma}}\left(\frac{\xi}{a}\right)^\ed \xiloc^\std \,.
\end{equation}
On the other hand, for the numerator we have to exercise more caution. Since the correlation function has the usual behavior of exponential decay in the local variables beyond $\xiloc$,
we can extend the integration region for these variables up to infinity in the limit $\mu\to 0$, thus leaving us with
\begin{align}
    \int_{\Omega_\xi} {\diff}^{\ed}\theta~{\diff}^\std\ff~ C(\pmb{\theta}, \pmb{\ff})
    &\approx\int_{\xinloc^\ed}\frac{\diff^\ed\theta}{\rv^\ed} \int_{\R^\std}\diff^\std\ff \sum_{\pmb{n}} \int\frac{\diff^\std k}{(2\pi)^\std} \frac{e^{i\pmb{n}\cdot\pmb{\theta}}e^{i\pmb{k}\cdot\pmb{\ff}}}{\alpha(\pmb{n})\sum_i k_i^2+\sum_{\ell=1}^\ed n_\ell^2/\rvp^2+b_{\pmb{n}}}\nonumber\\
    &=\frac{1}{\rv^\ed}\int_{\xinloc^\ed}\diff^\ed \theta\sum_{\pmb{n}}\frac{e^{i\pmb{n}\cdot\pmb{\theta}}}{\sum_{\ell=1}^\ed n_\ell^2/\rvp^2+b_{\pmb{n}}}\,
\end{align}
where the integral over local coordinates $\ff_i$ has been carried out first yielding again delta distributions $\delta(k_i)$ which simply eliminate all dependence on local variables upon integration over $k_i$.
Moreover, also the coupling $\alpha(\pmb{n})$ of these degrees of freedom to the local variables vanishes.
Therefore, the numerator of the quotient $Q$ is exactly the same as if there had been no local degrees of freedom at all, \eqref{eqn:sumovercontributions}. The result is that the local degrees of freedom contribute to $Q$ only in terms of the factor $\xiloc^{\std}$ in the denominator.

We treat the non-local degrees of freedom first in some more detail at finite compactness size $\cs$ and then in the large $\cs$ limit. 

\paragraph{Compact case.}

In the case of finite compactness scale $a$ we have seen in \eqref{eqn:correlationlengthcompact} that the correlation length of the geometric directions behave as $\xinloc^2 \sim a^2/24$ for ${\mu\to 0}$. 
Therefore, $\xinloc< a$
and we cannot extend the integration in the $Q$ parameter up to the whole of $G^r$. 
In order to make the computation more clear, let us split the sum into two contributions\footnote{This split is different in spirit from the one we performed in Section \ref{subsec:explicitexamplescorr}. Indeed, there it was performed in order to gain a better physical understanding of the properties of the geometric correlation function, while here it is just a computational aid.}: one where $\pmb{n}=\pmb{0}$ and one where $\pmb{n}\neq \pmb{0}$. We obtain
\begin{equation}
    \frac{1}{\rv^\ed}\int_{\xinloc^\ed}\diff^\ed \theta\sum_{\pmb{n}}\frac{e^{i\pmb{n}\cdot\pmb{\theta}}}{\sum_{\ell=1}^\ed n_\ell^2/\rvp^2+b_{\pmb{n}}}
    =\left(\frac{\xinloc}{a}\right)^\ed\frac{1}{b_{\pmb{0}}}+\frac{1}{\rv^\ed}\int_{\xinloc^\ed}\diff^\ed \theta\sum_{\pmb{n}\neq \pmb{0}}\frac{e^{i\pmb{n}\cdot\pmb{\theta}}}{\sum_{\ell=1}^\ed n_\ell^2/\rvp^2+b_{\pmb{n}}}\,.
\end{equation}
In the limit $\mu\to 0$, the first term diverges as $\mu^{-1}$, while the second one is finite because the series converges. Thus, it can be neglected with respect to the first one, and since $\xinloc/a=(24)^{-1/2}$, neglecting unimportant numerical factors, we can estimate the asymptotics of the numerator as 
$\sim b_{\pmb{0}}^{-1}\sim \mu^{-1}$ so that the $Q$-integral is
\begin{equation}\label{eq:Qscalarfield}
    Q\sim \frac{4\sum_\gamma\lambda_{\gamma}}{\mu^{2}}\xiloc^\std\sim \sum_\gamma\lambda_{\gamma}\xiloc^{4-\std}\,,
\end{equation}
and the critical dimension is $d_{\text{l,c}}=4$. Hence in the limit of compact Abelian Lie group $G$ one obtains the same result for the critical dimension as for a local statistical field theory on~$\mathbb{R}^\std$~\cite{zinn2007phase,Kopietz:2010zz}. This indicates that the mixed theory of local and non-local degrees of freedom becomes effectively local in this limit. Finally, notice that in~\eqref{eq:Qscalarfield} the strength of the minimal coupling of the local scalar degrees of freedom is absorbed in $\xiloc$ via~\eqref{eqn:mattercontribution} and would thus only enter through a simple pre-factor. This holds also in the ensuing scenario where we take the non-compact limit of $G$.

\paragraph{Non-compact case.}
In the case of non-compact Abelian Lie group $G$, instead, we will consider, as in the previous subsection, only the middle terms in the expansion of~\eqref{eqn:decompositioncorrelation}, whose typical contribution to the numerator $Q_N$ of the $Q$ parameter is still given by equation~\eqref{eqn:sumovercontributions}. Indeed, since for $\mu\to 0$ the correlation length $\xiloc$ diverges, we can extend the integration over the frame-variable part to infinity and essentially reduce the analysis to the one we have done above. As a consequence, we find again~\eqref{eqn:qnnoncompact} such that using
$\xinloc^2\sim \xiloc^2\sim \mu^{-1}\equiv \xi^2$, we have the result
\begin{equation}
    Q\sim \lambda_{\gamma}\frac{\xi^{4-\std-\gd(\rk-s_0)}}{\rv^{\gd(s_0-\rk)}}\,.
\end{equation}
The critical dimension for the combined system of non-local discrete geometric and local degrees of freedom is defined from $4=\std+\gd(\rk_c-s_0)$. We observe that the critical rank $r_c$ is reduced via the contribution stemming from the local degrees of freedom. Assuming their interpretation as (minimally coupled) scalar fields, this result demonstrates the impact that the inclusion of matter could have on the phase structure of more realistic quantum gravity models. Moreover, the result for the overall critical dimension suggests that the theory becomes effectively local in the non-compact limit of the Abelian Lie group $G$. Finally, it goes without saying that the more general expression~\eqref{eq:GinzburgQgeneralscaling} can be easily adapted to incorporate the impact of the local degrees of freedom yielding
\begin{equation}
    Q \sim\lambda_{\gamma}^{\frac{2}{\nv_\gamma-2}} \frac{\xi^{\frac{2 V_{\gamma}}{V_{\gamma-2}}-\std-\gd(\rk-s_0)}}{\rv^{\gd(s_0-\rk)}}\quad.
\end{equation}

As discussed at the end of Section \ref{subsec:ginzburgnoncompact}, the imposition of the gauge constraint leads to a shift of $r\to r-1$ or $s_0\to s_0+1$ in the previous expression.

\section{Discussion and conclusion}\label{sec:conclusion}

The main goal of this article was to examine, via Landau-Ginzburg mean-field theory, the phase structure and the phase transition between the broken and unbroken phase of the related global $\mathbb{Z}_2$-symmetry of different TGFT models. In particular, we considered models which include $\mathbb{R}$-valued local degrees of freedom. While the geometric degrees of freedom in TGFTs are non-local in the way they are coupled in the action, one motivation of the additional local degrees of freedom is that in TGFTs with a  quantum geometric interpretation, i.e. so-called group field theories (GFTs), they correspond to (free and massless, in the simplest models) minimally coupled (real) scalar fields, as confirmed via the simplicial gravity path integral expression of GFT Feynman amplitudes~\cite{Oriti:2016qtz,Li:2017uao,Gielen:2018fqv}. 

The first step of our work involved the proper set up of the Landau-Ginzburg method in the TGFT context and a careful discussion of the notion of correlation length for such systems of mixed local/non-local degrees of freedom. This allowed us to determine the conditions under which mean-field theory can be applied self-consistently through the Ginzburg criterion. In this way, the present work directly builds on previous ones on the application of mean-field theory to TGFT~\cite{Pithis:2018bw} as well as the application of the FRG methodology to the same class of quantum gravity models~\cite{Pithis:2020kio,Pithis:2020sxm}, and it extends this line of research by considering the influence of non-compact local directions of the domain of the group field. 

Assuming the matter interpretation of these local degrees of freedom, our work also marks an important first step towards the understanding of the phase structure and continuum limit of more realistic interacting quantum matter-quantum geometry systems. Notably, in spite of the assumptions of Landau-Ginzburg theory, most importantly the selection of the uniform field configuration, the truncation of the action functional and symmetries, our results agree with results from more involved renormalization studies of TGFTs~\cite{Carrozza:2016vsq,Benedetti:2015et,Benedetti:2016db,BenGeloun:2016kw,Pithis:2020kio}. This confirms the validity and the usefulness of the Landau-Ginzburg approach also in this TGFT context (the usefulness in usual QFT is of course well-established). 

In the following, we summarize as well as contextualize our main results and then comment on the limitations and possible extensions of our work.

For TGFT models on the \textit{non-compact} Abelian Lie group $G$ of dimension $d_G$, without gauge constraint but including $\std$ local $\mathbb{R}$-valued degrees of freedom we obtain for the critical case $2\frac{V_{\gamma}}{V_{\gamma}-2}=\std+\gd(\rk-s_0)$ wherein $r$ is the rank of the group field, $V_\gamma$ denotes the valency of the interaction and $s_0$ is the minimum number of zero-modes produced by it. For instance, $s_0=0$ holds if the interaction consists of disconnected melons, $s_0=1$ applies for one of connected melonic type, $s_0=r/2$ for necklace type and $s_0=r-1$ for simplicial interaction. Hence, the mean-field critical behavior is effectively that of a local scalar field theory on $\mathbb{R}^{\std+(r-s_0)d_G}$. 
As far as the non-local geometric degrees of freedom are concerned, this result was expected from the FRG analysis of cyclic-melonic TGFTs on $\text{U}(1)^r$ in~\cite{Pithis:2020kio,Pithis:2020sxm}. 

Importantly, our results show that melonic interactions drive the critical behavior of models characterized by different types of interactions (including melonic ones), which is again consistent with the literature~\cite{Bonzom:2011zz,Gurau:2011tj,Bonzom:2012hw,Gurau:2013cbh,Carrozza:2013oiy,Carrozza:2016vsq}. Finally, we note that when the gauge constraint is imposed onto the TGFT field, the expression for the critical dimension can be easily obtained by setting either $r\to r-1$ or $s_0\to s_0+1$ therein. 

In contrast, when $G$ is \textit{compact}, the critical dimension is $d_{\text{l,c}}=2\frac{V_{\gamma}}{V_{\gamma}-2}$ and thus solely determined by the local degrees of freedom, while the contribution stemming from the non-local degrees of freedom is completely washed away. This indicates that the critical behavior of the system is effectively described by a local theory on $\mathbb{R}^\std$, since the non-local group degrees of freedom become zero-dimensional. Indeed, this is supported by the FRG analysis in~\cite{Pithis:2020kio,Pithis:2020sxm} where it was shown in detail that the effective dimension of a cyclic-melonic TGFT on $\text{U}(1)^r$ flows to zero in the infrared quintessentially due to isolated zero-modes in the spectrum on the compact group strongly lending support to the expectation that the same should be the case for any TGFT on compact domain. Hence, our results illustrate that for a second-order phase transition to a condensate phase to occur for TGFTs, the non-compactness of the overall domain of the group field is crucially required, whether achieved by a non-compact group manifold encoding geometric data or by the addition of matter degrees of freedom.

Commenting further on the impact of the local degrees of freedom, it is clear that their presence improves the validity of mean-field theory at criticality. Given that they can be interpreted, at least in fully quantum geometric TGFT models, as (free, massless) scalar matter degrees of freedom~\cite{Oriti:2016qtz,Li:2017uao,Gielen:2018fqv}, our results indicate the important effect that matter degrees of freedom can have for the phase structure of discrete quantum gravity models. 

Interestingly, the strength of the coupling of the scalar degrees of freedom to geometric ones enters the Ginzburg $Q$-parameter only through a simple pre-factor absorbed in the definition of the correlation length, meaning that it has no influence on the value of the critical dimension. In light of this result, it would be interesting to go beyond the current setup and to investigate the impact of massive, interacting and non-minimally coupled local matter degrees of freedom onto the critical behavior of interacting quantum matter-quantum geometry systems. 

Notice, however, that our setup does not allow to address the question of how the critical behavior of the scalar fields themselves is altered by propagating on the effective geometry which emerges at criticality. This would correspond to a much more involved and totally different analysis left for future exploration. The difficulty comes again from the fact that we are considering critical behaviour, correlation length and phase transitions {\it of spacetime itself} to the extent in which it is captured by the TGFT models we considered, and not of QFT systems {\it on spacetime}, like usual scalar field theories. A spatiotemporal and directly physical translation of our employed notions and of our results in terms of physical correlations and observables requires more control over the emergent spacetime geometry and continuum fields. This will have to be crucial future work.

It is particularly important to emphasize this distinction between physical and \lq internal\rq~TGFT quantities, because our analysis, regarding the second pre-geometric type of quantities, is largely based on the physical intuition and expertise gained in usual, spacetime-based physics. Though this is arguably the most natural (and possibly only) way to proceed at this stage, when these models (and in particular their continuum approximation and effective macroscopic physics) are better understood, one may realize that a reassessment of our definitions and results is mandatory (see also~\cite{Oriti:2013jga}).

Further potential extensions of the method presented here could involve the relaxation of one of its main assumptions which is the projection onto uniform field configurations and to study the behavior of fluctuations around non-uniform minimizers~\cite{Fairbairn:2007sv,Girelli:2009yz,Girelli:2010ct,Livine:2011yb,BenGeloun:2018eoe} of TGFT models, which thus would retain non-local geometric information already in the structure of the non-trivial vacua and not merely in the fluctuations thereon. Considering non-uniform mean-field configurations also in the local variables could be crucial to connect to cosmological applications of GFTs~\cite{Gielen:2016dss,Oriti:2016acw,Pithis:2019tvp}, where the non-trivial dependence of the mean-field solution on one local variable allows for a relational interpretation of the evolution of geometry with respect to matter degrees of freedom~\cite{Oriti:2016qtz,Gielen:2018fqv,Marchetti:2020umh}. 
The impact of fluctuations in that context has already been studied~\cite{Gielen:2019kae,Marchetti:2020qsq,Gielen:2021vdd}, but so far only by assuming negligible interactions.

Even more compelling will be the mean-field analysis of the phase structure of quantum geometric Lorentzian TGFT models subject to closure and simplicity constraints (needed in order to capture fully geometric degrees of freedom and, in a continuum approximation, gravitational physics) together with local scalar matter, see for instance the model discussed in~\cite{Jercher:2021bie}. Encouragingly, for a gauge-invariant toy-model on $\text{SL}(2,\mathbb{R})$ (without local degrees of freedom) it has already been shown in~\cite{Pithis:2018bw,Pithis:2019mlv} that mean-field theory is sufficient to depict the transition between a broken and unbroken phase. Building on all of these results, the mean-field perspective on more realistic models will be addressed in forthcoming work~\cite{Marchetti:2021xyz}.

Since Landau-Ginzburg mean-field theory only gives an accurate account of the phase structure above the critical dimension, it is important to investigate the critical properties of the models considered here, in particular below the critical dimension, by means of non-perturbative RG techniques which allow to take into account fluctuations on all scales. A particularly effective way to do so is by means of the FRG~\cite{Berges:2002ga,Kopietz:2010zz,Dupuis:2020fhh}. We could anticipate from our results that an FRG analysis of such minimally coupled quantum matter-quantum geometry systems will find that above the critical dimension the Gaussian fixed point describes the phase transition with mean-field exponents. Below the critical dimension, we expect to find a rather intricate non-Gaussian fixed point structure similar either to that of local scalar field theories~\cite{Berges:2002ga,Codello:2012ec,Codello:2014yfa} (in particular when $G$ is compact) or to that of \enquote{standard} TGFTs not coupled to scalar matter, as found e.g. in~\cite{Benedetti:2015et,BenGeloun:2015ej,Benedetti:2016db,BenGeloun:2016kw,Pithis:2020kio,Pithis:2020sxm} (in particular when $G$ is non-compact). 

Ultimately, all these steps can also be seen as a necessary preparation to study the fixed-point structure and thus the continuum limit of full-blown GFT models for $4d$ Lorentzian quantum gravity including matter degrees of freedom using the FRG methodology. In perspective, this would also allow to compare the results obtained in these candidate fundamental and pre-geometric quantum gravity models with results on the phase configuration of interacting matter-geometry systems obtained instead from the continuum point of view, as in the asymptotic-safety program for quantum gravity~\cite{Percacci:2017fkn,Eichhorn:2018yfc,Eichhorn:2020mte,Bonanno:2020bil}.

\subsection*{Acknowledgments}
The authors thank D. Benedetti for discussions.

D. Oriti and A. Pithis acknowledge funding from DFG research grants OR432/3-1 and OR432/4-1.
The work of A. Pithis leading to this publication was also supported by the PRIME programme of the German Academic Exchange Service (DAAD) with funds from the German Federal Ministry of Education and Research (BMBF).
The work of J. Th\"{u}rigen was funded by the Deutsche Forschungsgemeinschaft (DFG, German Research Foundation) in two ways,
primarily under the author's project number 418838388 and
furthermore under Germany's Excellence Strategy EXC 2044--390685587, Mathematics M\"unster: Dynamics–Geometry–Structure. L. Marchetti thanks
the University of Pisa and the INFN (section of Pisa) for financial support, and the Ludwig Maximilians-Universit\"at (LMU) Munich for the hospitality.

\appendix

\section{Useful integrals}\label{app:integrals}
In this Appendix, we provide some useful formulas for the explicit computation of correlation functions. 

In general, the typical integral needed to obtain the correlation function in coordinate space is given by
\begin{equation}
    I_D(\mu)\equiv \int\diff^Dp\frac{e^{i\mathbf{p}\cdot\mathbf{x}}}{\vert\mathbf{p}\vert^2+\mu}=\int\diff\Omega_D\int_0^\infty\diff p\, p^{D-1}\frac{e^{i\mathbf{p}\cdot\mathbf{x}}}{p^2+\mu}\,,
\end{equation}
where $p^2\equiv \vert\mathbf{p}\vert^2\equiv \sum_{i=1}^Dp_i^2$, $D$ is non-zero positive integer, $D\in \mathbb{N}^+$, and $\Omega_D$ is the angular measure on the Euclidean $D$-dimensional space. Defining $r^2\equiv \vert\mathbf{x}\vert^2\equiv \sum_{i=1}^Dx_i^2$ and $\mathbf{p}\cdot\mathbf{x}\equiv pr\cos\theta$, the above integral can be rewritten as~\cite{Hong_Hao_2010}
\begin{equation}
    I_D(\mu)=\frac{2\pi^{\frac{D-1}{2}}}{\Gamma(\frac{D-1}{2})}\int_0^\pi\sin^{D-2}\theta\diff\theta\int_0^\infty\diff p\, p^{D-1}\frac{e^{ipr\cos\theta}}{p^2+\mu}\,.
\end{equation}
The angular integral can be easily performed, following~\cite{Hong_Hao_2010}
\begin{equation}
    \int_0^\pi\sin^{D-2}\theta\diff\theta\,e^{ipr\cos\theta}=\sqrt{\pi}\left(\frac{2}{pr}\right)^{\frac{D-2}{2}}\Gamma\left(\frac{D-1}{2}\right)J_{\frac{D-2}{2}}(pr)\,,
\end{equation}
where $J_\alpha(z)$ is the Bessel function of the first kind. Combining these last two equations and changing the variable to $q\equiv pr$, we obtain
\begin{equation}
    I_D(\mu)=\frac{2^D\pi^{D/2}}{r^{D-2}}\int_0^\infty\diff q\frac{q^{D/2}}{q^2+\mu r^2}J_{\frac{D-2}{2}}(q)\,.
\end{equation}
The explicit value of the integral now depends crucially on the sign of $\mu$. Let us discuss these two cases separately.
\begin{description}
\item[\underline{$\mu>0$}:] When $\mu>0$ the integrand is regular over the whole domain of integration and through the residue theorem one finds that
\begin{equation}
    I_D(\mu)=\frac{2^D\pi^{D/2}}{r^{D-2}}(\mu r^2)^{\frac{D-2}{4}}K_{\frac{D-2}{2}}(\sqrt{\mu} r)=2^D\pi^{D/2}(\mu r^{-2})^{\frac{D-2}{4}}K_{\frac{D-2}{2}}(\sqrt{\mu} r)\,,
\end{equation}
where $K_\alpha(z)$ is the modified Bessel function of the second kind, the asymptotic behavior of which for large $z$ and for $\vert\text{arg}(z)\vert <3\pi/2$ is
\begin{equation}\label{eqn:asymptoticexpmodbessel}
    K_\alpha(z)\sim \sqrt{\frac{\pi}{2z}}e^{-z}\left(1+\mathcal{O}(z^{-1})\right)
\end{equation}
Hence, for large values of $r$ the correlation function is exponentially suppressed. The scale $\sqrt{\mu}$ sets the scale of exponential suppression and thus can immediately be interpreted as the correlation length of the system.
\item[\underline{$\mu<0$}:] When $\mu<0$, the integrand is singular on the domain of integration. One can still perform the integration by employing the Feynman prescription, so that $q^2+\mu r^2\to q^2+(\mu -i\epsilon)r^2$, with $\epsilon\in \mathbb{R}$ being set to zero after integration. The result is
\begin{equation}
    I_D(\mu)=2^D\pi^{D/2}(-\vert\mu \vert r^{-2})^{\frac{D-2}{4}}K_{\frac{D-2}{2}}(i\sqrt{\vert \mu\vert}r)\,.
\end{equation}
From the asymptotic expansion~\eqref{eqn:asymptoticexpmodbessel}, we see therefore, that besides unimportant proportionality factor, the behavior of this integral for large $r$ is
\begin{equation}
    I_D(\mu)\sim 2^{D-1/2}\pi^{D/2+1}(-\vert\mu \vert r^{-2})^{\frac{D-2}{4}} \frac{e^{-i\sqrt{\vert \mu \vert}r}}{(\sqrt{\vert \mu\vert} r)^{1/2}}\,.
\end{equation}
The function is therefore oscillating and suppressed by $r^{D-3/2}$ at large $r$.
\end{description}

\bibliographystyle{jhep}
\bibliography{references.bib} 

\end{document}

%% file: fdiagrams.tex
\definecolor{jred}{rgb}{0.8,0,0}
\definecolor{jgreen}{rgb}{0,0.7,0}
\definecolor{jblue}{rgb}{0,0,0.8}

\usetikzlibrary{arrows,patterns,shapes,positioning,fit}
\tikzstyle{c} =	[coordinate]
\tikzstyle{vc} = [circle, draw=black, line width=1pt, inner sep=0pt, minimum size=2mm]
\tikzstyle{v} =  [circle, draw=black, line width=.2pt, fill=black, inner sep=0pt, minimum size=1.5mm]
\tikzstyle{vb} =[circle, draw=black, line width=.1pt, fill=jred, inner sep=0pt, minimum size=1mm]
\tikzstyle{vh} =[circle, draw=black, line width=.1pt, fill=jblue, inner sep=0pt, minimum size=1.5mm]
\tikzstyle{vs} =[coordinate]
\tikzstyle{e} =	[draw=jred,line width=1pt]
\tikzstyle{eb}= [draw=jgreen,line width=1pt]
\tikzstyle{es}= [dashed, line width=1pt]
\tikzstyle{ev}= [draw=jgreen,line width=1pt]
\tikzstyle{f} = [line width=0.01pt,dotted,fill=blue, fill opacity=.1]
\tikzstyle{cs}= [ellipse, draw=black, line width=.1pt, fill=jred, inner sep=1pt, minimum size=2mm]

\newcommand{\vertexexstranded}{
\begin{tikzpicture}[x=6ex,y=6ex,baseline={([yshift=-.59ex]current bounding box.center)}] 
    \node [vs]	(12)	at (1,.1)	{};
    \node [vs]	(13)	at (1,0)	{};
    \node [vs]	(14)	at (1,-.1)	{};    
    \node [vs]	(21)	at (.1,1)	{};
    \node [vs]	(24)	at (0,1)	{};
    \node [vs]	(23)	at (-.1,1)	{};    
    \node [vs]	(32)	at (-1,.1)	{};
    \node [vs]	(31)	at (-1,0)	{};
    \node [vs]	(34)	at (-1,-.1)	{};    
    \node [vs]	(41)	at (.1,-1)	{};
    \node [vs]	(42)	at (0,-1)	{};
    \node [vs]	(43)	at (-.1,-1)	{};
    \node [vc, fit=(13) (31)] {};
    \path
    \foreach \i/\j in {14/41,21/12,32/23,43/34}{
    (\i) edge [ev,bend right=40] node [vh] {} (\j)
    }
    \foreach \i/\j in {13/42, 31/24}{
    (\i) edge [ev,bend right=60] node [vh] {} (\j)
    };
    \node [cs,fit=(12) (14)] {};
    \node [cs,fit=(21) (23)] {};
    \node [cs,fit=(32) (34)] {};
    \node [cs,fit=(41) (43)] {};
\end{tikzpicture}
}

\newcommand{\vertexexgraph}{
\begin{tikzpicture}[x=6ex,y=6ex,baseline={([yshift=-.59ex]current bounding box.center)}]    
\scriptsize{
    \node [vb, label=right:${1}$]	(2)	at (0,1)	{};
    \node [vb, label=below:${2}$]	(3)	at (-1,0)	{};
    \node [vb, label=left:${3}$]	(4)	at (0,-1)	{};
    \node [vb,label=above:${4}$]	(1)	at (1,0)	{};
    \node [v, fill=black, minimum size =3mm]   (0) at (0,0)    {};
    \path
    (1) edge [ev] (2)
    (3) edge [ev] (4)
    \foreach \i in {1,2,3,4}{
    (\i) edge [e] (0)
    }
    \foreach \i/\j in {1/4,4/1,2/3,3/2}{
    (\i) edge [ev,bend right=20] node [c] {} (\j)
    };
}
\end{tikzpicture}
}

\newcommand{\vertexexeom}[3]{
\begin{tikzpicture}[x=4ex,y=4ex,baseline={([yshift=-.59ex]current bounding box.center)}]    
\scriptsize{
    \node [vb, label=above:${#1}$]	(2)	at (0,1)	{};
    \node [vb, label=below:${#2}$]	(3)	at (-1,0)	{};
    \node [vb, label=below:${#3}$]	(4)	at (0,-1)	{};
    \node [c]   (a) at (.7,1)   {};
    \node [c]   (b) at (.7,-.8) {};
    \node [c]   (c) at (.7,-1.2){};
    \node [v, fill=black]   (0) at (0,0)    {};
    \path
    (a) edge [ev] (2)
    (3) edge [ev] (4)
    (b) edge [ev] (4)
    (c) edge [ev] (4)
    \foreach \i in {2,3,4}{
    (\i) edge [e] (0)
    }
    \foreach \i/\j in {2/3,3/2}{
    (\i) edge [ev,bend right=20] node [c] {} (\j)
    };
}
\end{tikzpicture}
}

\newcommand{\vertexexvg}{
\begin{tikzpicture}[rotate=45,x=1ex,y=1ex,baseline={([yshift=-.59ex]current bounding box.center)}]    
    \node [vb]	(1)	at (1,0)	{};
    \node [vb]	(2)	at (0,1)	{};
    \node [vb]	(3)	at (-1,0)	{};
    \node [vb]	(4)	at (0,-1)	{};
    \path
    (1) edge [ev] (2)
    (3) edge [ev] (4)
    \foreach \i/\j in {1/4,4/1,2/3,3/2}{
    (\i) edge [ev,bend right=40] node [c] {} (\j)
    };
\end{tikzpicture}
}

\newcommand{\cvtvg}{
\begin{tikzpicture}[x=1ex,y=1ex,baseline={([yshift=-.59ex]current bounding box.center)}] 
\node [vb]	(1)	at (0,1)	{};
\node [vb]	(2)	at (0,-1)	{};
\path
\foreach \i/\j in {1/2,2/1}{
   (\i)	edge [ev,bend left=15]	(\j)
   (\i)	edge [ev,bend left=40]	(\j)
   };
\end{tikzpicture}
}

\newcommand{\cvfvg}{
\begin{tikzpicture}[x=1ex,y=1ex,baseline={([yshift=-.59ex]current bounding box.center)}] 
\scriptsize{
\node [vb, label=above:]	(b1)	at (-1,1)	{};
\node [vb, label=below:]	(w1)	at (-1,-1)	{};
\node [vb, label=below:]	(b2)	at (1,-1)	{};
\node [vb, label=above:]	(w2)	at (1,1)	{};
\path
\foreach \i/\j in {1/2}{
   	(w\i) edge [ev]  node [c, label=below:$\ell$] {} (b\j)
	(b\i) edge [ev]   (w\j)
	}
\foreach \i in {1,2}{
	(w\i)   edge [ev]				(b\i)
	(w\i)	edge [ev,bend left=30]	(b\i)
	(w\i)	edge [ev,bend right=30]	(b\i)
	};
}
\end{tikzpicture}
}

\newcommand{\cvfnvg}{
\begin{tikzpicture}[x=1ex,y=1ex,baseline={([yshift=-.59ex]current bounding box.center)}] 
\scriptsize{
\node [vb, label=above:]	(b1)	at (-1,1)	{};
\node [vb, label=below:]	(w1)	at (-1,-1)	{};
\node [vb, label=below:]	(b2)	at (1,-1)	{};
\node [vb, label=above:]	(w2)	at (1,1)	{};
\path
\foreach \i/\j in {1/2,2/1}{
   	(w\i) edge [ev,bend left=30]  (b\j)
   	(w\i) edge [ev,bend right=30] (b\j)
	}
\foreach \i in {1,2}{
	(w\i)	edge [ev,bend left=30]  (b\i)
	(w\i)	edge [ev,bend right=30]	(b\i)
	}

\foreach \i in {w1,b1,w2,b2}{
	};
}
\end{tikzpicture}
}

\newcommand{\cvf}{
\begin{tikzpicture}[x=2ex,y=2ex,baseline={([yshift=-.59ex]current bounding box.center)}] 
\scriptsize{
\node [v]	(1)	at (0,0)	{};
\node [vb, label=above:]	(b1)	at (-1,1)	{};
\node [vb, label=below:]	(w1)	at (-1,-1)	{};
\node [vb, label=below:]	(b2)	at (1,-1)	{};
\node [vb, label=above:]	(w2)	at (1,1)	{};
\path
\foreach \i/\j in {1/2}{
   	(w\i) edge [ev]   (b\j)
	(b\i) edge [ev]  node [c, label=above:$\ell$] {} (w\j)
	}
\foreach \i in {1,2}{
	(w\i)   edge 	[ev]				(b\i)
	(w\i)	edge [ev,bend left=30]	(b\i)
	(w\i)	edge [ev,bend right=30]	(b\i)
	}
\foreach \i in {w1,b1,w2,b2}{
   	(\i) edge [e] (1)
	};
}
\end{tikzpicture}
}

\newcommand{\cvft}{
\begin{tikzpicture}[x=2ex,y=2ex,baseline={([yshift=-.59ex]current bounding box.center)}] 
\node [v]	(v)	at (0,0) 	{};
\node [vb]	(1)	at (-1,1)	{};
\node [vb]	(2)	at (-1,-1)	{};
\node [vb]	(3)	at (1,-1)	{};
\node [vb]	(4)	at (1,1)	{};
\path
\foreach \i in {1,2,3,4}{
    (\i) edge [e] (v)
    }
\foreach \i/\j in {1/2,2/1,3/4,4/3}{
   (\i)	edge [ev,bend left=15]	(\j)
   (\i)	edge [ev,bend left=40]	(\j)
   };
\end{tikzpicture}
}

\newcommand{\cvfn}{
\begin{tikzpicture}[x=2ex,y=2ex,baseline={([yshift=-.59ex]current bounding box.center)}] 
\scriptsize{
\node [v]	(1)	at (0,0)	{};
\node [vb, label=above:]	(b1)	at (-1,1)	{};
\node [vb, label=below:]	(w1)	at (-1,-1)	{};
\node [vb, label=below:]	(b2)	at (1,-1)	{};
\node [vb, label=above:]	(w2)	at (1,1)	{};
\path
\foreach \i/\j in {1/2,2/1}{
   	(w\i) edge [ev,bend left=30]  (b\j)
   	(w\i) edge [ev,bend right=30] (b\j)
	}
\foreach \i in {1,2}{
	(w\i)	edge [ev,bend left=30]  (b\i)
	(w\i)	edge [ev,bend right=30]	(b\i)
	}
\foreach \i in {w1,b1,w2,b2}{
   	(\i) edge [e] (1)
	};
}
\end{tikzpicture}
}

\newcommand{\cvfsvg}{
\begin{tikzpicture}[x=1ex,y=1ex,baseline={([yshift=-.59ex]current bounding box.center)}] 
\scriptsize{
\foreach \i in {72,144,216,288,360}{
\begin{scope}[rotate=\i]
\node [vb]	(\i)	at (0,1.6)	{};
\end{scope}
}
\path
\foreach \i in {72,144,216,288,360}{
    \foreach \j in {72,144,216,288,360}{
   	    (\i) edge [ev]  (\j)
	    }
	};
}
\end{tikzpicture}
}

\newcommand{\cvfs}{
\begin{tikzpicture}[x=2ex,y=2ex,baseline={([yshift=-.59ex]current bounding box.center)}] 
\scriptsize{
\node [v]	(0)	at (0,0)	{};
\foreach \i in {72,144,216,288,360}{
\begin{scope}[rotate=\i]
\node [vb]	(\i)	at (0,1.6)	{};
\end{scope}
}
\path
\foreach \i in {72,144,216,288,360}{
    \foreach \j in {72,144,216,288,360}{
   	    (\i) edge [ev]  (\j)
	    }
    }
\foreach \i in {72,144,216,288,360}{
  	(\i) edge [e] (0)
	};
}
\end{tikzpicture}
}